\documentclass[fleqn,usenatbib]{mnras}

\usepackage{newtxtext,newtxmath}
\usepackage{comment}
\usepackage{graphicx}
\usepackage{adjustbox}
\usepackage{placeins}

\usepackage[T1]{fontenc}

\DeclareRobustCommand{\VAN}[3]{#2}
\let\VANthebibliography\thebibliography
\def\thebibliography{\DeclareRobustCommand{\VAN}[3]{##3}\VANthebibliography}

  \DeclareGraphicsExtensions{.pdf,.png}

\usepackage{graphicx}	
\usepackage{amsmath}	

\usepackage{amssymb}	

\title[TACOS: \emph{TESS} AM~CVn Outbursts Survey]{TACOS: \emph{TESS} AM~CVn Outbursts Survey}

\author[M. Pichardo Marcano et al.]{
Manuel Pichardo Marcano,$^{1}$\thanks{E-mail: manuel.pichardo-marcano@ttu.edu}
 Liliana E. Rivera Sandoval,$^{2}$
 Thomas J. Maccarone$^{1}$
and Simone Scaringi$^{3}$
\\
$^{1}$Department of Physics and Astronomy, Texas Tech University, Lubbock, TX 79409, USA\\
$^{2}$Department of Physics, University of Alberta, CCIS 4-183, Edmonton, AB, T6G 2E1, Canada\\
$^{3}$Centre for Extragalactic Astronomy, Department of Physics, University of Durham, South Road, Durham, DH1 3LE, UK
}

\date{Accepted XXX. Received YYY; in original form ZZZ}

\pubyear{2015}

\begin{document}
\label{firstpage}
\pagerange{\pageref{firstpage}--\pageref{lastpage}}
\maketitle

\begin{abstract}

Using \emph{TESS} we are doing a systematic study of outbursting AM~CVn systems to place some limits on the current outbursts models. We present the \emph{TESS} light curve (LC) for 9 AM~CVns showing both superoutbursts (SOs) and  normal outbursts (NOs). The continuous coverage of the outbursts with \emph{TESS} allows us to place stringent limits on the duration and structures of the SOs and the NOs. We present evidence that in at least some of the systems enhanced mass transfer (EMT) has to be taken into account to explain the observed LC of the SOs and rebrighthening phase after the SOs. For others, the colour evolution from simultaneous observations in $g$ and $r$ with ZTF differs from previously reported colour evolution of longer period AM~CVns where EMT is responsible for the SO. We also find that due to the lack of sufficiently high cadence coverage the duration of many systems might have been overestimated in previous ground-based surveys. We report the SO duration for 6 AM~CVns. We also found that precursors are a common feature of SOs in AM~CVns and are seen in the LC of 5 of the 6 reported SOs. Finally the 10-minute and 2-minute cadence LCs from \emph{TESS} also allowed us to find two new candidate orbital periods of AM~CVns, both of which are in reasonably good agreement with the predictions for their periods based on their past outburst histories.

\end{abstract}




\section{Introduction}

AM Canum Venaticorum (AM CVn) stars are a class of rare compact binaries in which a white dwarf (WD) accretes hydrogen-poor and helium-rich material from a degenerate (or semi-degenerate) companion. These objects are characterized by their strong helium  
spectral lines; their lack of (or very weak) hydrogen features in their spectra; and by their short orbital periods ($<$70 min). For a review see \cite{Solheim2010}.

These rare accretors are of special interest as significant persistent gravitational wave sources. AM~CVns are expected to contribute to an unresolved gravitational wave background and the shortest period and/or nearest of them can serve as verification sources for the future space mission \emph{LISA} \citep[e.g.][]{Nelemans2004lisa}.

The new generation of large sky surveys is steadily discovering AM~CVn candidates. Spectroscopic surveys like the Sloan Digital Sky Survey \citep{2000AJ....120.1579Y} and wide-area photometric variability surveys like the All-Sky Automated Survey for Supernovae \citep[ASAS-SN;][]{ASASSN2014}, and the Zwicky Transient Facility \citep[ZTF;][]{ZTF2019Masci} have increased the number of AM~CVns from about 10 \citep[][]{RoelofsSDSS} to $\sim 70$ \citep[][]{vanroestel2021}.  
Many of these sources are discovered by these new photometric surveys as optical transients via their outburst events. 

The thermal viscous disk instability model (DIM) is one approach for describing these outbursts, and was originally developed for explaining  dwarf nova (DN) outbursts from normal cataclysmic variables (CVs) and soft X-ray transient outbursts in X-ray binaries \citep[e.g.][]{Meyer1981DIM,Smak1982,Cannizzo1982ApJ...260L..83C}. In this model, the outbursts are triggered by an instability in the accretion disk due to a sudden change in the `viscosity'\footnote{The inflow is not driven by molecular viscosity, but rather probably by the magnetorotational instability \citep[][]{BalbusHawley}.} parameter $\alpha$ at temperatures at which hydrogen (in DNe) or helium (in AM~CVns) is partially ionized. For a review on the DIM for DNe see \cite{HameuryDIMReview2020AdSpR}.

Many of the known AM~CVns show superoutbursts (SOs) in addition to their normal outbursts (NOs). The bulk of these superoutbursts is thought to be similar to the ones shown by the (mostly) short period dwarf novae systems called SU~UMa stars \citep{Warner1995SOSUUMAAp&SS.226..187W}. During the superoutbursts, these systems show periodic brightness modulations, with periods close to their orbital periods, called `superhumps'.

One model to explain the periods of these superhumps and their associations with the superoutbursts is the thermal-tidal instability model (TTI) developed by \cite{Osaki1989SOModelPASJ...41.1005O}. 
In this model, the superoutbursts in SU~UMa systems are produced when the outer edge of the accretion disk reaches the 3:1 resonance radius due to the accumulation of mass and momentum by normal outbursts, in which the mass transfer rate is larger than the mass accretion rate. The observed period of the superhumps is identified with the beat period between the orbital period of the binary and the precession period of the now temporally eccentric disk. 
This model has been disputed by some authors. For example, \cite{Smak2009AcA....59...89S,Smak2009AcA....59..103S,Smak2009AcA....59..121S,Smak2009AcA....59..419S}, argues against the eccentric disk model and favors an enhanced mass transfer model (EMT) where the modulations are due to the dissipation of the kinetic energy of the accretion stream due to variable irradiation from the hot  white  dwarf and the superoutbursts are due to a change in the rate of mass transfer from the donor star \citep{LasotaEMTmodel1995A&A...302L..29L,BuatEMTmodel2002A&A...386..891B}. More recently, the TTI has been called into question by \cite{Cannizzo2012ApJPrecursor} by studying the quiescent intervals between normal outbursts. Subsequent studies with CVs have tried to place constraints on the models with high cadence data \citep{Osaki2013KeplerV15}, arguing again for the TTI to explain the superoubursts in SU~UMas.

The disk instability model has been applied to AM~CVns as well \citep{Smak1983AcA....33..333SHelium,Cannizzo1984NaturDIM,Tsugawa1997PASJ...49...75T,ElKhoury2000A&A,Menou2002,Lasota2008UltraCompact,Kotko2012} and observational studies have tried to constrain how well this model describes the SOs in these systems.  For example, \cite{Levitan2015longterm} studied the long-term behaviour of these superoutbursts in AM CVns and found empirical relations (1) between the orbital period and the superoutburst recurrence time, (2) the outburst amplitude and the orbital period and (3) between the orbital period and the duration of the superoutburst. The latter relationship has been explained by \cite{CannizzoNelemansDIM2015} and \cite{ Cannizzo2019Duration} in the context of a DIM model. \cite{CannizzoNelemansDIM2015} predicted the relation should be much flatter than that found by \cite{Levitan2015longterm} and later confirmed by \cite{Cannizzo2019Duration} by exploring this relation excluding upper limits and adding observations of AM~CVns systems with periods longer than 40 min such as SDSS~J1411+4812 \citep{2019RSM} and CRTS~J0450 \citep{Levitan2015longterm}.
Two recent papers have empirically established that at least a fraction of the superoutbursts in AM~CVn systems must have origin in mass transfer instabilities \citep[][]{RiveraSandovalYearlong2020ApJ...900L..37R,RiveraSandoval60minOutburst2020arXiv201210356R}. In both of these long-period systems, SDSS~J080710+485259 and SDSS~J113732+405458 (from now on SDSS~0807 and SDSS~1137 with $P_{orb}\sim 53$ and $60$ min, respectively), the outbursts are dramatically longer (more than a year duration) than any reasonable estimates for the outburst timescales in disk instability models \citep[see also ][for the co-discovery of the long duration of this outburst]{Wong2021RNAAS...5....3S}. Furthermore, the outbursts in these long-period systems have been shown to become redder when brighter. That colour evolution is a problem for any class of ionization driven disk instability model because the disk always remains cooler than the central white dwarf and never reaches a colour blue enough \footnote{Though in the case of the binary SDSS~J080710+485259 a substantial heating of the disk could have occurred at late times, turning it bluer.} to indicate that helium should have become ionized, strongly favouring a mass transfer instability mechanism. The existence of these outbursts in such long-period systems is also a problem because most versions of the ionization instability model predict that AM~CVn binaries at such long periods (and hence extremely small mass transfer rates) should be persistently in the low state and never outburst. Though, the orbital period where the transition from unstable to cold and stable disks occurs is not yet well established.  

AM~CVns, much like the DNe, are expected not only to show SOs but to show NOs. These NOs are very common in DNe, but their durations, which are of order the viscous timescale through the accretion disk, have traditionally made them hard to detect in  AM~CVn systems due to the AM CVns' ultrashort periods. The lack of continuous coverage at short cadence has made it difficult to differentiate between NOs and SOs. A few systems that have been studied with targeted long ground-based photometric campaigns have been found to show some candidate NOs \citep[e.g.][]{KatoNormalOCRBOO2000MNRAS.315..140K,Levitan2011PTF0719,DuffyTESSKLDRA2021}, but many of these remain unconfirmed. Due to their short durations and inadequate cadence in the wide-field ground-based transient surveys, the NOs can be missed entirely, or merged or confused with a SO, and their durations and peak brightnesses cannot be adequately measured.

However, new space missions like \emph{Kepler} and the
\textit{Transiting Exoplanet Survey Satellite} \emph{(TESS)}, offer unique long continuous coverage to study a large variety of binaries, including AM~CVns \citep[e.g.][]{Fontaine2011SDSS1908,Green2018SDSS1351,Kupfer2015SDSS1908,DuffyTESSKLDRA2021}. 

Here we report results of the first dedicated study of the outbursting behaviour of AM~CVn systems with \emph{TESS}. We present and discuss detailed \emph{TESS} light curves (LCs) of 9 AM~CVns, which show NOs and SOs. Furthermore, the short cadence of \emph{TESS} also allowed us to determine new periods for two AM~CVns. These identified periods  are close to the predicted orbital periods by their superoutburst properties \citep{Levitan2015longterm}.

\section{Observations}
\subsection{TESS data}
In this work, we exploit the unique capabilities of \emph{TESS} \citep{TESS2014SPIE} which allow us to collect continuous, high cadence observations of AM~CVns in outburst\footnote{During quiescence most AM~CVns are faint and therefore most of the times below the \emph{TESS} detection limit.}. This has permitted us to obtain the most detailed short-term LCs of these binaries so far. Using four CCD cameras, \emph{TESS} obtains continuous optical images of a rectangular field of $24 \deg \times\ 90 \deg$ for 27.4 days at a cadence at least as fast as 30 minutes. 

\emph{TESS} was launched in 2018 with the intention of mapping the entire sky over a two year period but due to its success, the mission has been extended. During the first and third years of the mission (July 2018-July 2019 and July 2020-July 2021), the southern ecliptic hemisphere was observed, and during the second year (July 2019-July 2020), the northern hemisphere was monitored. During these first nominal two years, 26 sectors were observed. From sector 27, the extended mission started with 10-minute full-frame images instead of 30-minute ones. 

Our survey consists of \emph{TESS} data obtained under programs G022237 and G03180 (P.I. Rivera Sandoval). It makes use of the full-frame images (FFI), which have 30-minute cadence  earlier in the mission and 10-minute cadence for the most recent data.  Selected targets have been observed with a 2-minute cadence. AM~CVns in both hemispheres were included in the programs and observed by \emph{TESS}. Here, results are presented only for those systems that showed statistically significant outburst activity.

\subsection{ZTF Public Survey Data}

A subset of these outbursts have substantial numbers of detections in ground-based data, which can be used to follow the colours of the outbursts and to help calibrate the \emph{TESS} magnitude scale for the individual objects, which is normally difficult because the large \emph{TESS} pixels typically include many stars. We thus also use public data from the ZTF in the $g$ and $r$ filters. Simultaneous observations of these surveys with the ones of \emph{TESS} allowed us to estimate the amplitudes of the outbursts. 

We also study AM~CVn colour evolution using the $r$ and $g$ magnitudes from ZTF. For each of the $r$ measurements, we pick the closest $g$ measurement  within an interval of 5 hrs.

\section{Data Analysis}
\subsection{TESS light curves of AM~CVns}
We use the Python package for \emph{Kepler} and \emph{TESS} data analysis \texttt{Lightkurve} \citep[v1,][]{lcsoftware2018} to create the LCs from the FFI. To extract our LCs, we select pixels around the coordinates of the target, and `empty' pixels (with no Gaia sources) as our background, and which have a flux that is $0.0000001 \sigma$ below the median flux. We also used the package \texttt{eleanor} \citep{EleanorFeinstein2019} to extract systematics-corrected LCs and compare them to our LCs from the FFI using the median background removal method. 

For the targets with 2-minute cadence, we use the Pre-search Data Conditioning Simple Aperture Photometry flux (PDCSAP), in which long-term trends have been removed using the so-called Co-trending Basis Vectors \citep[CBVs, ][]{KeplerLCMethod2012}.  These have been produced by the \emph{TESS} Science Processing Operations Center \citep[SPOC, see][]{TESSSPOC2016SPIE}. The obtained LCs are in flux units of e/s and not calibrated to standard magnitudes. For the analysis we use the LCs normalized to the median flux.

\subsection{Timing Analysis}
 
 Thanks to the fast readout of \emph{TESS} we can search for periodicity in the LCs. The CCDs of the spacecraft are read at 2-second intervals that are then grouped to produce 2-minute, 10-minutes or 30-minutes cadence depending on the target and the sector\footnote{https://tess.mit.edu/science/}. 
 
For the sources with 2-minute and 10-minute cadence, we search for periodicity during both the normal outbursts and the superoutbursts. However, before performing a search for periodicity we detrend the events from residual background fluctuations or artefacts. For that purpose, we fitted two polynomials to the normal outbursts, one for the rise and one for the decay phase (see Fig.~\ref{fig:plateaua14cc} bottom panel). For the superoutbursts we also fitted a polynomial but this only to the `plateau' phase (see Fig.~\ref{fig:plateaua14cc} top panel and section~\ref{subsection:SOAnal}) because it was the longest and most affected phase. We detrended the LCs using the fitting routines from Astropy and the affiliated package \emph{specutils} \citep{astropy:2018,specutilnicholas_earl_2021_4603801}. For the plateau phase we used a straight-line fit and to detrend the background we fitted a Chebyshev1D polynomial of degree 3.

For the timing analysis we used the Lomb-Scargle \citep{Lomb1976,Scargle1982} and Phase Dispersion minimization \citep[PDM,][]{Stellingwerf1978PDM} techniques as implemented also in the Astropy and in the PyAstronomy packages \citep{pyastronomypackage}.
To calculate the false alarm probability (FAP), which is an estimate of the significance of the minimum against the hypothesis of random noise for the PDM, \cite{pdmbeta97} showed that the PDM statistic follows a $\beta$ distribution and thus the FAP can be calculated as:

$$\text{FAP} = 1-\left [ 1 - \beta\left ( \frac{N-M}{2},\frac{M-1}{2},\frac{(N-M)\times \Theta}{N-1}\right )\right ]^m$$

\noindent where N is the total number of data points, M is the number of bins, $\Theta$ is the value of the PDM theta statistic, and $m$ is the effective number of independent frequencies. The number of independent frequencies can be estimated in a number of ways \citep[e.g.,][]{Nemec1985,HorneLombFreq1986,Cumming2004}. Here, we follow the conservative prescription given in \cite{GuidePeriodSearch2003}  and choose $m= \min(N, N_f,\Delta f \Delta T)$ where $N_f$ is the number of frequencies, $\Delta f$ is the frequency range and $\Delta T$ is the time span of observations. For the Lomb-Scargle we use the bootstrap method to approximate probability based on bootstrap re-samplings of the input data.  This approach is not robust to red noise variations, but the candidate periodicities we find have large numbers of cycles, and hence red noise tests are not required.

\subsection{Outburst Analysis}\label{subsection:SOAnal}

In this paper 'normal outbursts' and superoutburst of the sources with orbital periods ~22-31 minutes have been identified when there is an increase in the flux ($>3 \sigma$ from the mean noise level) of the LC. Whenever available, the SOs and NOs were confirmed with simultaneous ground-based data available from the ZTF Public Survey.

For the case of normal outbursts we define these as having a rather sharp shape and lasting less than 3 days from the beginning to the end, based on predictions of the DIM model \citep{Smak1999AcA....49..391S} and from current limits of previous ground-based monitoring of these systems \citep[e.g.][]{Levitan2015longterm,DuffyTESSKLDRA2021}. For the case of the SOs we also define them as an increase in the flux of the LC but with timescales longer than $\sim 2$ days and having a distinctive `plateau' phase. We note that there is a 1-to-1 correspondence between the outbursts that are longer than 2 days and the outbursts that have distinct plateau phases.

\begin{figure}
	\includegraphics[width= 1.0 \columnwidth]{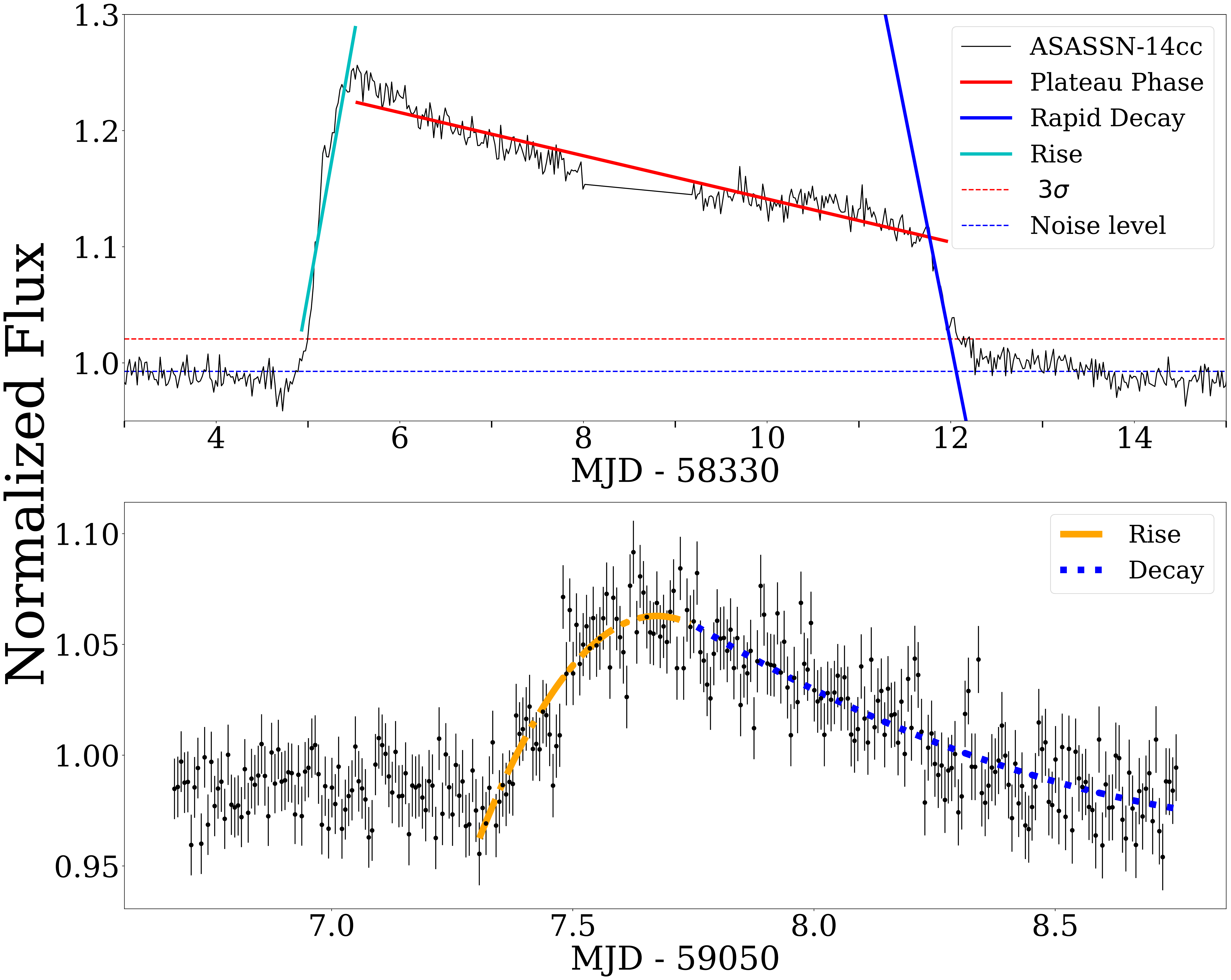}
    \caption{\emph{Top Panel}: Zoom-in of the superoutburst of ASASSN-14cc in Sector 1 showing how we determine the duration of the SO for our sources. The SO is divided into 3 different phases: the rise (shown in solid cyan), the plateau (shown in solid red line), and the decay (shown in solid blue). The beginning of the SO is defined as the moment in which the flux is $3 \sigma$ above the average noise level (dashed red line), the noise level is shown in the dashed blue line. We define the beginning of the SO 'plateau' as the maximum of the LC and the end of the phase as where the two solid lines meet. \emph{Bottom Panel}: A further zoom-in on the outburst of ASASSN-14cc in Sector 27 with the fit for the rise and decay overplot to the \emph{TESS} LC. }
    \label{fig:plateaua14cc}
\end{figure}

In order to characterize the superoutbursts, we have divided them into 3 main different phases: the rise, the plateau and the decay. This is illustrated for the source ASASSN-14cc in fig. \ref{fig:plateaua14cc}.  The rise phase goes from the beginning of the SO (with the start defined as the first data point $3\sigma$ above the baseline flux level) to the maximum. 
The plateau phase extends from the time of maximum brightness to the beginning of the fast decay. The end of the fast decay is when the SO reached the original noise level. The break time between the plateau and the fast decay is determined by fitting two straight lines, one to the plateau phase and one to the fast decay phase. The data points to fit the lines to both regions are determined through visual inspections.

Due to the big pixel size of \emph{TESS} (21 arcseconds), in crowded regions there is possible contamination by nearby sources. This can dilute the variability, but for the sources analyzed here, contamination does not change the conclusions as the shape of the SOs presented in this manuscript is clearly distinguishable above the noise level. 

Some of our targets are near the detection limit of \emph{TESS}. For fainter stars around \emph{TESS} magnitude, $T_{mag} = 16$ (600 - 1000 nm), the photometric precision is about 1\% \citep{TESShandbook2019ESS.....433312V}\footnote{https://heasarc.gsfc.nasa.gov/docs/tess/observing-technical.html}. This means that some of the possible fainter outbursts might get confused by the background. Here we have focused on the SOs and NOs detected by \emph{TESS} with a brightness increase of more than $3\sigma$ above the baseline flux level. Many of these NOs and SOs are also confirmed via simultaneous ZTF data.

\section{Results and discussion}

In total, 9 sources showed outbursting activity during our campaigns.   Given the diverse behaviour of the outbursts in AM~CVns, we present the results for each source along with the discussion regarding the possible nature of their variable behaviour. 


\begin{table*}
\centering
\caption{\label{tab:table-name}Table of the sources that showed either a superoutburst (SO) or a normal outburst (NO) as seen by \emph{TESS}. For the orbital period, (p) implies the predicted orbital period based on the outburst properties \citep{Levitan2015longterm} and (sh) means superhump period, which is typically close to the orbital period and longer by a few percent. Magnitude range taken from \citet[][]{RamsayCatalog} and \citet[][]{vanroestel2021}.}
\label{tab:table-detection}
\resizebox{\linewidth}{!}{%
\begin{tabular}{|c|c|c|c|c|c|c|} 
\hline
\textbf{Source} & \textbf{Period (min)} & \textbf{$\Delta$ mag} &\textbf{TESS Sectors [ISO date]} &\textbf{No. SOs and outbursts} & \textbf{SO duration (day)} & \textbf{Comments}                                                                                                              \\ 
\hline
ASASSN-14cc     & 22.5 (sh) & 16 – 20 (V)  &  \begin{tabular}[c]{@{}c@{}}1 [2018-07-25 to 2018-08-22] \\ 27-28 [2020-07-05 to 2020-08-25 ]\end{tabular}     & 2 SOs and 3 NOs       & 7.2                  & \begin{tabular}[c]{@{}c@{}c@{}}No precursor detected in both SO      \\ SO1: 58334.99 to  58342.06 MJD \\ SO2: 59067.46 to 59074.65 MJD     \end{tabular} \\

                                                                                          \\ 
\hline
PTF1 J2219+3135  & 28 (candidate)  & 16.2–20.6 &  16 [2019-09-12 to 2019-10-06 ]    & 1 NO                  & $-$                          & Periodicity in normal outburst                                                                                                 \\ 
\hline
V803 Cen        & 26.6  & 12.8–17.0  &  11 [2019-04-23 to 2019-05-20]          & at least 4 NOs         & $-$                          & \begin{tabular}[c]{@{}c@{}}Very activate many fainter outbursts \\and some possibly confused with the background\end{tabular}  \\ 
\hline
PTF1 J0719+4858 & 26.8   & 15.8–19.4  &  20 [2019-12-25  to 2020-01-20]    & 1 SO, 3 echo outbursts       & 5.2            & \begin{tabular}[c]{@{}c@{}c@{}}Precursor and at variable echo outbursts \\ with increase brightness after the plateau    \\
SO: 58852.93 - 58858.12 MJD

\end{tabular} \\ 
\hline
KL Dra & 25 & 16.0–19.6 &  14-26 [2019-07-18 to 2020-07-04]   & 5 SOs and many echos/normal outbursts & 6 & \begin{tabular}[c]{@{}c@{}c@{}c@{}} 
\emph{TESS} LC first reported by \cite{DuffyTESSKLDRA2021} \\ 
SO1: 58789.17-58796.19 \\ SO2: 58852.45-58858.50 \\ SO3:58916.74-58922.059

\end{tabular}\\ 

\hline
CP Eri          & 28.4 & 16.2–20.2  &  \begin{tabular}[c]{@{}c@{}}4 [2018-10-19 to 2018-11-14] \\ 31 [2020-10-22 to 2020-11-16]\end{tabular}           & 1 SO and 1 echo O     & 5.8              & \begin{tabular}[c]{@{}c@{}} Short precursor and at least one echo outburst \\     SO: 58427.97 - 58433.82 MJD\end{tabular}\\ 
\hline
SDSS J1043+5632 & 28.5 (p)  & 17.0–20.3 &  21 [2020-01-21  to 2020-02-18]  & 1 SO and 8 echo outbursts     & 6        & \begin{tabular}[c]{@{}c@{}}Precursor and many echo outbursts.\\

SO: 58878.97 - 58885.013
 MJD\end{tabular}\\ 

\hline
Gaia 16all       & 31.1 (candidate) & 16.2 – 20.6 (G) &  \begin{tabular}[c]{@{}c@{}}1-13 [2018-07-25 to 2019-07-17] \\ 27-33 [2020-07-05 to 2021-01-13]\end{tabular}      & 3 SOs 1 NO             & 5.4                      & \begin{tabular}[c]{@{}c@{}c@{}}
Limit on recurrence time and many echo outbursts  \\
SO1: 58442.510 - 58447.887 MJD\\
SO2: 58642.011 -  58647.556 MJD\\
SO3: > 3.7 MJD\\
\end{tabular}\\ 
\hline

ZTF18abihypg & $31.2$ (p) & 15-19.9 &  \begin{tabular}[c]{@{}c@{}}16-17 [2019-09-12 to 2019-11-02]  \\ 24 [2020-04-16 to 2020-05-12] \end{tabular}  &2 NOs & $-$ & ZTF LC reported by \cite{vanroestel2021} \\

\hline
\end{tabular}
}
\end{table*}

\subsection{ASASSN-14cc: A system with normal outbursts and superoutbursts with no precursors}

ASASSN-14cc was discovered in outburst by ASAS-SN and later confirmed by \cite{Kato2015} as an AM~CVn, who also performed follow-up during four superoutbursts and measured the period of the superhumps to be 22.5 min.

\begin{figure}
	\includegraphics[width= 1.0 \columnwidth]{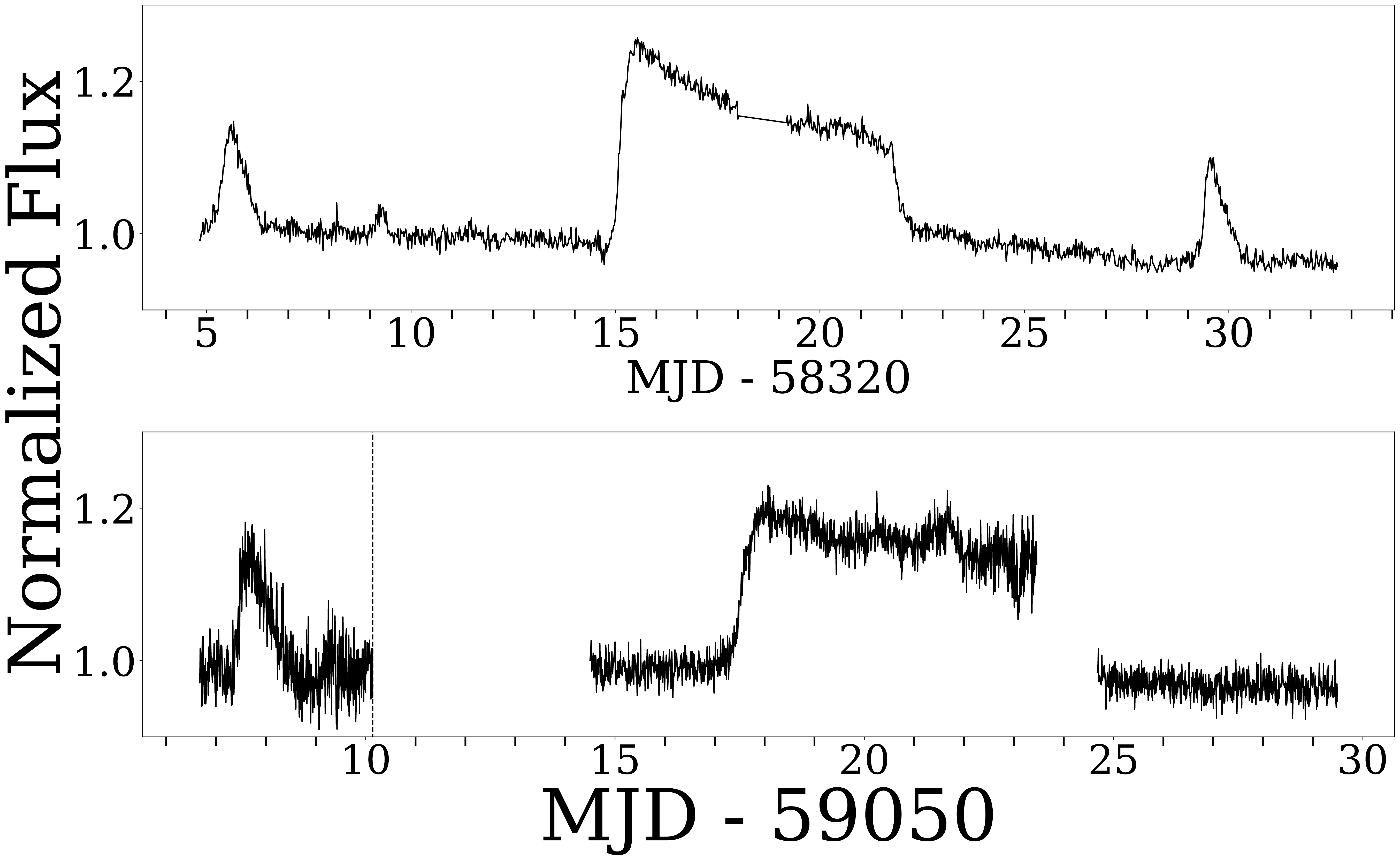}
    \caption{LC from sector 1 at 30-minutes cadence (top panel) and sector 27-28 at 10-minutes (bottom panel) for ASASSN-14cc. The LC in sector 1 (top panel) shows no precursor and a normal outburst before and after the SO. The SO of sectors 28 also lacks a precursor before the SO and an outburst ~10 days before the SO. The end of the LC of the SO in sector 28 is contaminated by the background but the end of the 'plateau' phase and beginning of the  fast decay phase is still clear. The dashed black line represents the division between sector 27 and 28.  }
    \label{fig:asssn14ccSector1}
\end{figure}

ASASSN-14cc was observed by \textit{TESS} in three sectors (1, 27, 28). Sector 1 was observed from 58324.8 MJD (2018-07-25) to 58352.8 MJD (2018-08-22) with 30-minutes cadence (see Fig.\ref{fig:asssn14ccSector1}). Sectors 27 and 28 were observed with a 10-minutes cadence from 59035.8 MJD (2020-07-05) to 59086.6 MJD (2020-08-25, Fig. \ref{fig:asssn14ccSector1}). The final part of Sector 27 shows a normal outburst, while Sector 28 shows a superoutburst. The fast cadence of Sector 28 allowed us to check for periodic variability during the plateau phase of the superoutburst. We find a peak in the periodogram at 22.5 min that agrees with the superhump period for this source found by \cite{Kato2015} (see fig.~\ref{fig:LombScargle14ccappendix} in the Appendix).

The LCs of ASASSN-14cc (Sector 1 and 28) where SOs where recorded do not show a precursor. The fact that a precursor is not observed in either SO indicates that such behaviour is not due to artefacts, but rather, it is intrinsic to the source. This behaviour is contrary to that seen in SOs of other WD systems studied at short cadence, including some DNe observed with \emph{Kepler} \citep{Cannizzo2012ApJPrecursor} and recently the AM~CVn system KL~Dra, also observed with \emph{TESS} \citep{DuffyTESSKLDRA2021}, where a precursor was detected. However, some SOs in SU UMa-Type DNe have also shown a lack of precursor, which has been revealed using high cadence photometry from \emph{Kepler}  \citep[e.g.][]{2013PASJ...65...97K}. 

In a refined version of the TTI model, \cite{OsakiMeyer2003A&A...401..325O} explained that the lack or presence of a precursor depends on whether the accretion disk passes the 3:1 resonance radius and reaches the tidal truncation disk or not. According to that model, if the tidal truncation radius is reached, the matter that is damped at that radius causes a gradual decay without a precursor. In this way, SU~UMas with large mass ratios can show SOs with and without precursors. Under this model the sizes of the accretion disks would be different for systems with and without a precursor, being larger in the cases where a precursor is not observed. \cite{UemuraNoprecurso2005A&A...432..261U} proposed that the behaviour of $\dot P_{sh}/P_{sh}$,  where $P_{sh}$ is the superhump period, is  related to the amount of gas around and beyond the 3:1 resonance radius. That ratio is positive for longer time if the amount of mass is larger. The transition from positive to negative $\dot P_{sh}/P_{sh}$ would be explained by the depletion of gas.

The lack of a precursor during the superoutburst of ASASSN-14cc suggests that an analogous situation might be occurring. Unfortunately, despite the fact that we have detected superhumps during the superoutburst, we are unable to study their evolution given our current coverage. Higher signal-to-noise and shorter cadence would be needed to test whether the scenario proposed by \cite{UemuraNoprecurso2005A&A...432..261U} is valid or not in AM~CVns. 

Using the method described in the data analysis section, we place some limits on the duration of the superoutburst of ASASSN-14cc. In Sector 1, the superoutburst lasted 7.1 days and the `plateau' phase lasted 6.2 days. These numbers were similar for Sector 28 in which the superoutburst and the `plateau' phase lasted 7.2 and 6.2 days, respectively. This conflicts with previously reported values for the duration of the superoutbursts for this source. For example, recently, \cite{DuffyTESSKLDRA2021} using ground-based data reported a superoutburst duration of $14\pm 2.2$ days. Considering the cadence and coverage of that data set \citep[also reported previously by][]{Kato2015}, and the coverage and cadence of our \emph{TESS} data, it is possible that the change in the duration of the SO is intrinsic to the source, perhaps due to a change in the size of the accretion disk, or due to changes in the mass transfer rate.

\subsection{PTF1 J2219+3135: an AM~CVn with a possible 28 minutes period}

\begin{figure}
	\includegraphics[width= 1.0 \columnwidth]{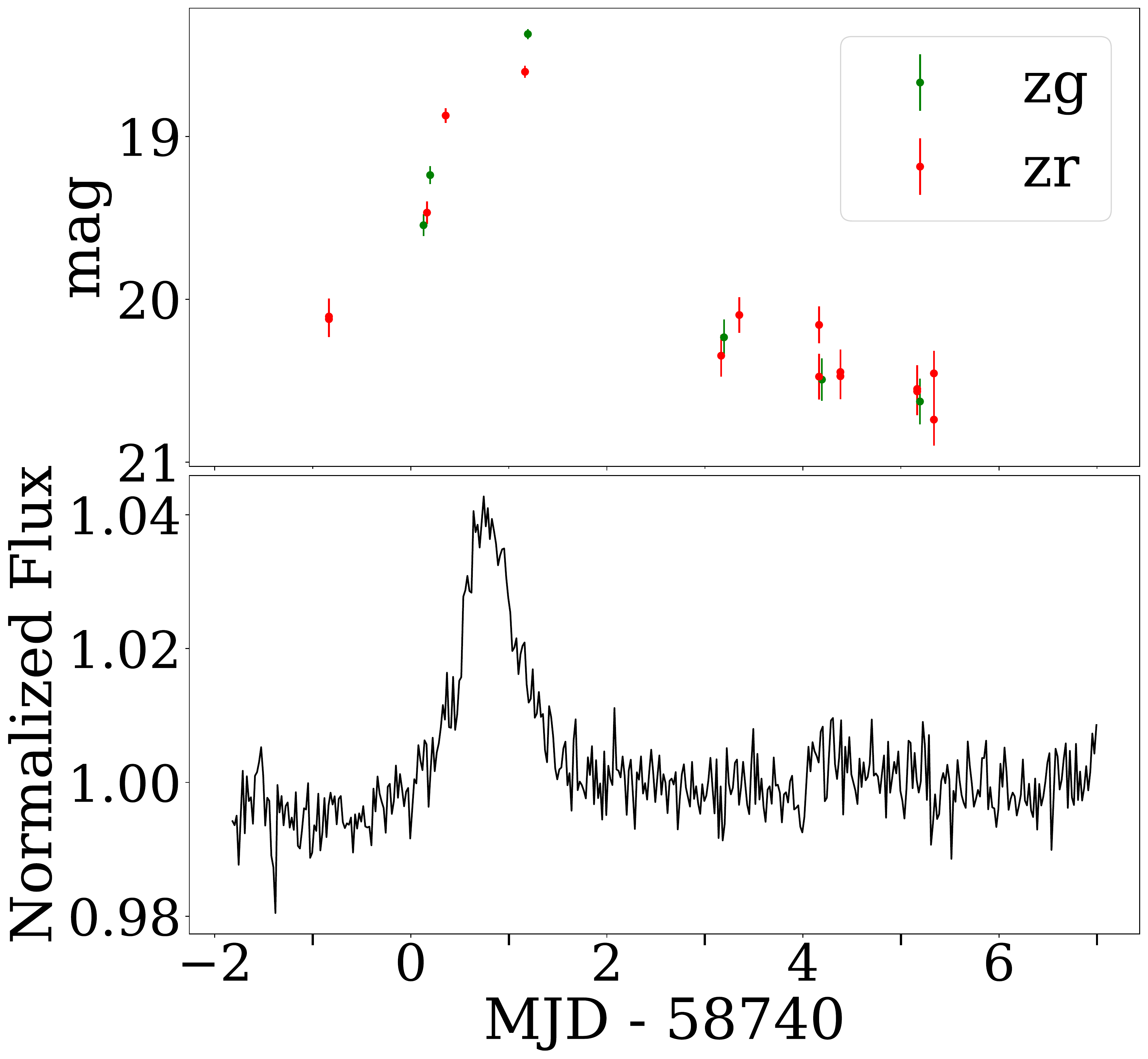}
    \caption{LC for PTF1 J2219+3135 from sector 16 from ZTF (top panel) and \emph{TESS} (bottom panel). The normal outburst seen with \emph{TESS} is also seen in ZTF data as $\sim 1.5$ mag outburst. This source was observed at 2-minute cadence }
    \label{fig:PTF2219ZTF}
\end{figure}

PTF1 J2219+3135 was discovered by \cite{Levitan2013PTF1}. It was observed by \emph{TESS} in sector 16 from 58738.2 MJD (2019-09-12) to 58762.8 MJD (2019-10-06). This source was observed at 2-minutes cadence and showed an outburst (fig.~\ref{fig:PTF2219ZTF}).

The period of this source is not known. Based on the outburst properties, it has been predicted to be 26 minutes \citep{Levitan2015longterm}. With the 2-minute cadence data we search for periodicity during the normal outburst.

\begin{figure}
	\includegraphics[width= 1.0 \columnwidth]{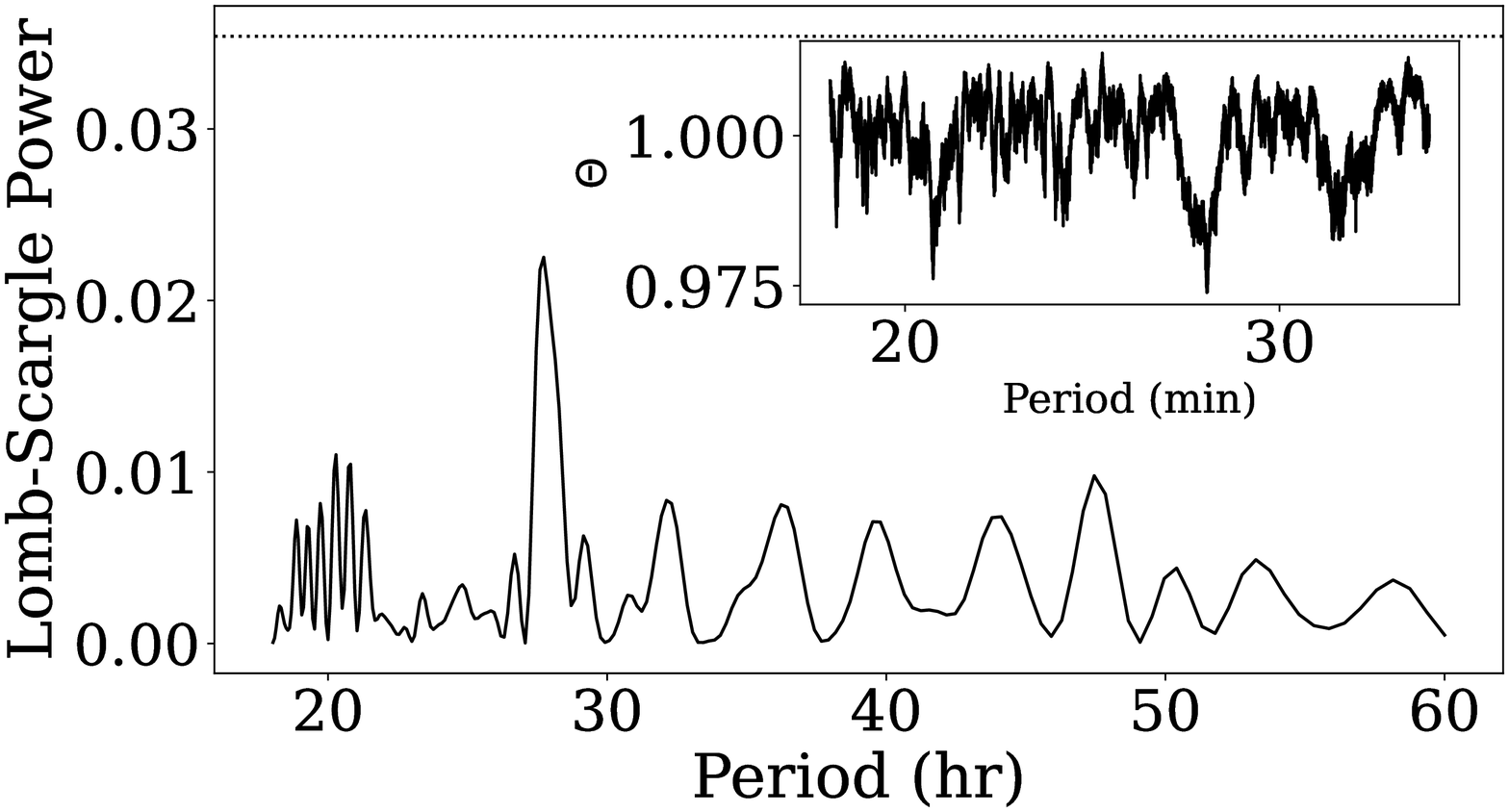}
    \caption{Lomb-Scargle periodogram for the decay part of the outburst of PTF1J2219+3135. The peak corresponds to a period of 27.7 minutes. This period is also recovered using PDM (inset).}
    \label{fig:PTF2219PDM}
\end{figure}

The periodogram shows a peak at 27.7 minutes (fig.~\ref{fig:PTF2219PDM}). We also performed PDM analysis and found that the $28.1$ min value
 minimizes the variance of the LC (inset Fig.~\ref{fig:PTF2219PDM}). This gives a FAP of 0.12 for the period found by PDM. Because the statistical significance for the detection is marginal, further confirmation via spectroscopy is required, but the agreement between the candidate photometric period and the period predicted by the outburst properties is strongly suggestive and should be taken as good motivation to design a search optimized for periods in the 25-30 minute range.

If the period is real, it might represent the orbital period of the source;  we cannot, however, discard the possibility that this periodicity is the superhump period since superhumps have been observed during normal outbursts of SU~UMa systems \citep{Imada2012PASJ...64L...5I,Kato2012PASJ...64...21K}. If this period is confirmed as the superhump period, this would make PTF1J2219+3135 the first AM~CVn showing superhumps during normal outbursts.

Recently \cite{DuffyTESSKLDRA2021} reported a ground-based LC for this source but the authors stress that the sampling was not optimal. 
Their data suggest that normal outbursts occur for this source, but the authors mentioned they could not "show convincing evidence of normal outbursts in PTF1 J2219+3135" or measure the duration of these putative normal outbursts. Here we confirm their suspicion and show the LC of an outburst with a duration of roughly ~1 day. This adds to the few AM~CVns that convincingly show normal outbursts and not only SOs.

\subsection{V803 Cen: normal outbursts with variable amplitude}

The first high-speed photometry for this object was reported by \cite{V8031987MNRAS.227..347O}, who classified it as an AM~CVn and found a periodicity of 26.85 minutes (1611 seconds) for the source, likely a superhump period. A spectroscopic study by \cite{V803Roelofs2007MNRAS.379..176R} later revealed the orbital period to be $1596.2 \pm 1.2$ s.

In 2000, \cite{V803Patterson2000PASP..112..625P} concluded that V830 had several distinct states: 1) a high state that lasts for 3-12 days with $V = 12.7-13.3$, 2) a low state with $V = 17$ and duration of at least 10-30 days, and 3) a `cycling state' in which the star varied rapidly between $V=13.4-14.5$ with a period of $22\pm 1$ hr.

Regarding the cycling  state, \cite{V803Patterson2000PASP..112..625P} concluded that the brightenings in the cycling state followed the Kukarkin-Parenago \citep{Kukarkin1969gcvs.book.....K,KPRelationAntipova1987Ap&SS.131..453A} relation for DNe and novae, and concluded that these were `normal' outbursts similar to the ones found in dwarf novae CVs. Contrary, \cite{KatoStandStillV8032001IBVS.5091....1K} equated this `cycling' state as similar to the standstills that are observed in Z~Camelopardalis (Z~Cam) DNe systems, and classified this system (along with CR Boo) as peculiar helium ER~UMa-type stars. Later, \cite{Kotko2012} using the \cite{V803Patterson2000PASP..112..625P} cycling light curve tried to model these systems and suggested the existence of superoutbursting Z~Cam type AM~CVns.

\begin{figure}
	\includegraphics[width= 1.0 \columnwidth]{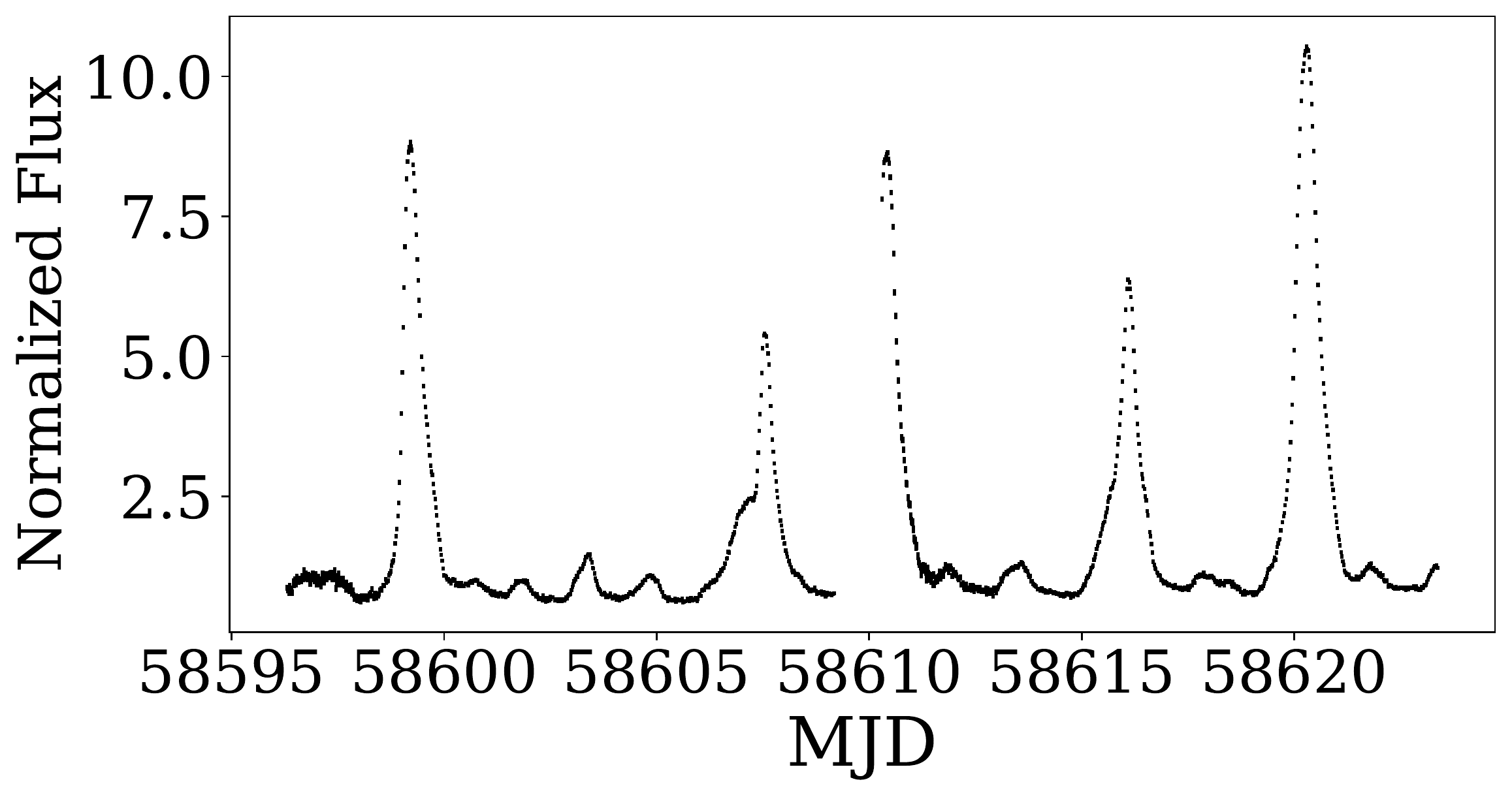}
    \caption{LC of V803 Cen in sector 12 at 30-minutes cadence showing continuous normal outburst of changing amplitudes.  }
    \label{fig:V803out}
\end{figure}

This source was monitored in sector 11 by \emph{TESS} from 2019-04-23 to 2019-05-20 at 30 minutes cadence. We do not observe any superoutbursts, but we do see some outbursts in the LC (Fig.~\ref{fig:V803out}). The recurrence time of superoutbursts for this source was found to be $\sim 77$ days by \cite{Kato2004PASJ...56S..89K}, therefore it is not surprising that we do not detect any in our 27 day monitoring with \emph{TESS}.

The LC of fig.~\ref{fig:V803out}, shows the detailed behaviour of the partially detected "supercycle" for this AM~CVn. During quiescence this object has a magnitude close to the \emph{TESS} detection limit. Thus, we are sensitive to all brightness increases, with errors in their duration of no more than a few hrs. From that figure one can also see that the behaviour of the normal outbursts in V803~Cen is very peculiar. A `high amplitude' outburst is followed by at least one very small amplitude outburst, which also has a shorter duration. One possible explanation is that in the case of the larger outbursts, the heating front reached the outer disk radius and the smaller outbursts are produced when the front does not reach the outer disk regions. This happens when the mass transfer rate is relatively low and the disk mass does not grow enough between two normal outbursts. We note that the relative shapes between the high and short amplitude outbursts resembles somehow to the "stunted outburst" phenomenology seen in some nova-like systems \citep[e.g.][]{Honeycutt1998AJ....115.2527H,Ramsay2016MNRAS.455.2772R}. However, we believe it is unlikely that this is more than a coincidence, given that the nova-like systems have significantly higher mean mass transfer rates, and longer orbital periods (and hence larger expected disk-crossing timescales), but further theoretical work should be done to determine more conclusively whether these phenomena might be related. 

The behaviour shown by V803~Cen heavily contrasts with the models by \cite{Kotko2012} for Z~Cam like AM~CVns, where the amplitude of the normal outbursts gradually increases as they are closer to the next SO. Under that model that behaviour is the result of combining mass transfer modulations with the EMT. The \emph{TESS} LC then argues against such a model for V803~Cen. Instead, it supports more a model that includes the DIM with a larger metallicity and no substantial change of $\alpha$ during the hot and cold states \citep[see e.g. Fig. 5 of ][]{Kotko2012}.

\subsection{PTF1 J0719+4858: evidence for enhanced mass transfer}

PTF1 J0719+4858 was discovered using data from the Palomar Transient Factory (PTF) by
\cite{Levitan2011PTF0719}. The same authors measured an orbital period of $26.77 \pm 0.02 $ minutes via spectroscopy. Then \cite{Han2021NewA...8701604H} found a superhump period of 1659.85(13) s (27.66 minutes). Due to its faintness ($g \sim 19$ in quiescent), the outburst behaviour of PTF1 J0719+4858  has not been very well studied. Recently, \cite{DuffyTESSKLDRA2021} reported on ground-based observations of the source where SOs were detected but they did not report the presence of NOs between SOs. 

This AM CVn star  was observed in \textit{TESS} sector 20 from 58842 MJD (2019-12-25) to 58868.3 MJD (2020-01-20). 
Unlike \cite{DuffyTESSKLDRA2021}, a small amplitude ($r \sim 18.4$ mags) outburst was detected on 58846.24 MJD which is confirmed by the ZTF data as shown in figure \ref{fig:ptf0719ZTFandTESSshaded}. Unfortunately, the ZTF data does not allow us to precisely determine the duration of the event due to the lack of coverage, while the \emph{TESS} data does not constrain the duration due to the faintness of the object, which rapidly felt below the detection limit. 
However, with the \emph{TESS} and ZTF data the recurrence time of ~10 days for normal outbursts reported for this binary by \cite{Levitan2011PTF0719} was not observed. Considering that \emph{TESS} only observed $\sim 10$ days before the start of the SO and some of the reported NOs by \cite{Levitan2011PTF0719} are close to the detection limits of \emph{TESS} ($g \sim 18.4 $), the recurrence might be hard to confirm without help from other surveys.

In addition to the NO, a SO was also observed. The SO shows a precursor and has a total duration of 5.19 d. Interestingly, after the end of the superoutburst the binary did not reach the original quiescence level.  Instead, a series of echo outbursts occurred. They had durations of 0.7, 2 and 5 days, respectively (fig. \ref{fig:ptf0719ZTFandTESSshaded}). The amplitude of the echo outbursts was also variable. The whole behaviour observed by \emph{TESS} was mirrored in the ZTF data. Due to 30-minute sampling for this source (larger than the expected superhump period), a period analysis was not performed. 

\cite{Kotko2012}, modelled the behaviour of PTF1~J0719+4858 as a helium SU~UMa star under a DIM with accretion-irradiation mass transfer rate increase.  While the LC obtained by these authors is not totally similar to the one observed by TESS, there are characteristics that are broadly consistent with expectations from the DIM with enhanced mass transfer. First, the presence of a precursor is typical of the EMT model. Second, there is an observed gradual increase in the flux (and hence magnitude) of the echo outbursts, especially evident between the first and the second one. This increase can be explained because during each echo outburst the disk gains more mass than it loses due to accretion. This leads to an accumulation of mass and the surface density of the disk then increases as well.  Furthermore, the duration of the echo outbursts increases. The fact that the binary does not reach the original quiescent level after the SO, but instead it stays stuck at a higher flux, indicates changes in the mass transfer rate \citep[see  e.g.][]{Hameuryrebright2021}. All these characteristics are typical of the EMT model. The observations of PTF1~J0719+4858 then argues for the need to account for EMT mechanisms in the DIM in AM~CVns with $P_{orb}$ shorter than 30 min. This is not surprising as such mechanisms have been shown to exist at longer periods \citep{RiveraSandovalYearlong2020ApJ...900L..37R,RiveraSandoval60minOutburst2020arXiv201210356R} where enhanced mass transfer and not disk instabilities are the main cause for the outbursts.

\subsubsection{The colour evolution of PTF1~J0719+4858}

\begin{figure}
	\includegraphics[width= 1.0 \columnwidth]{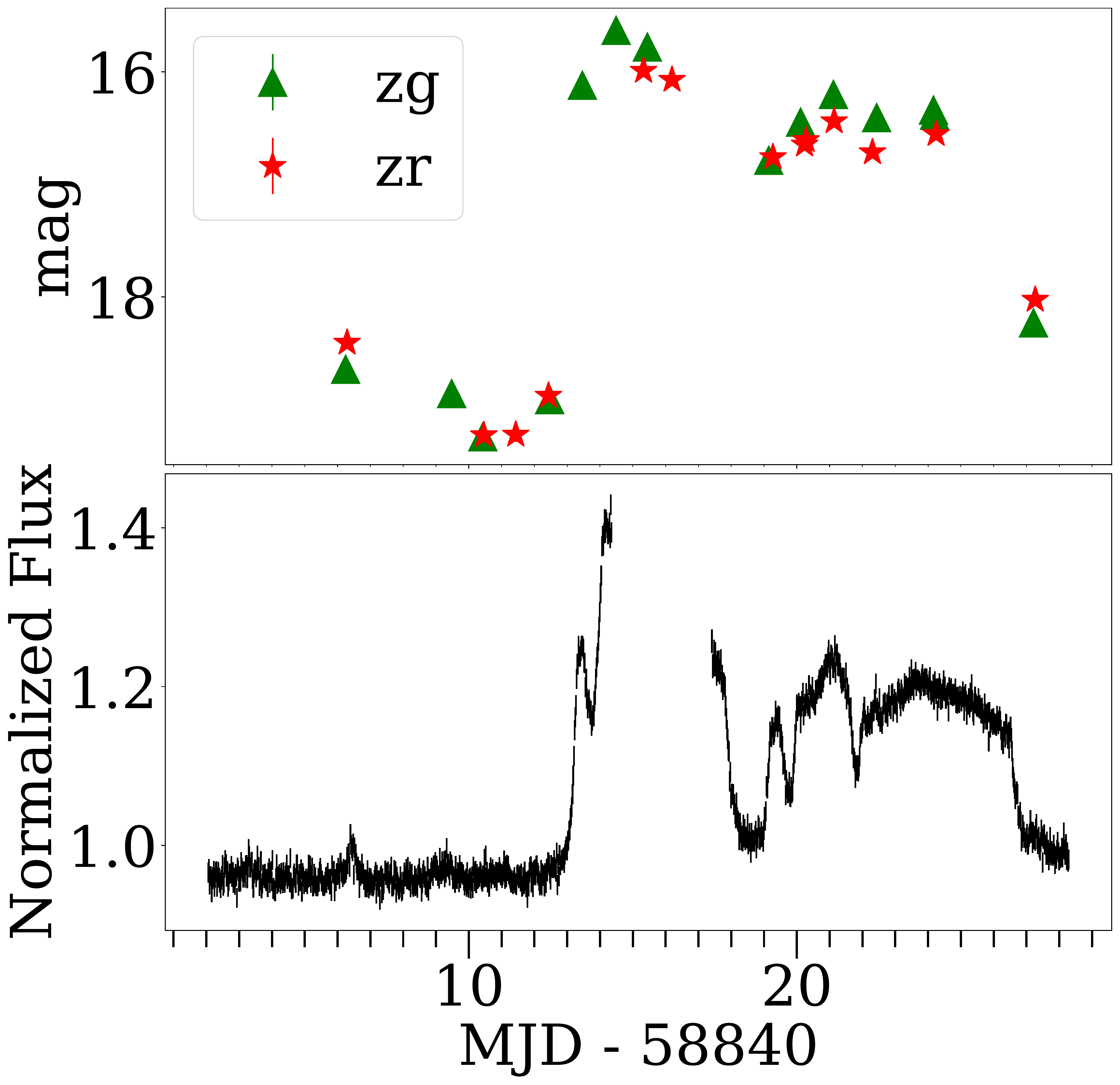}
    \caption{LC from ZTF (top) and \emph{TESS} (bottom) LC of PTF1 J0719+4858 from sector 20 at 30-minute cadence. The data from the middle of the plateau phase of the SO is contaminated with background and taken out for clarity. The LC shows a small outburst before the SO and at least 3 echo outbursts of increasing duration and variable brightness after the SO.}
    \label{fig:ptf0719ZTFandTESSshaded}
\end{figure}

\begin{figure}
	\includegraphics[width= 1.0 \columnwidth]{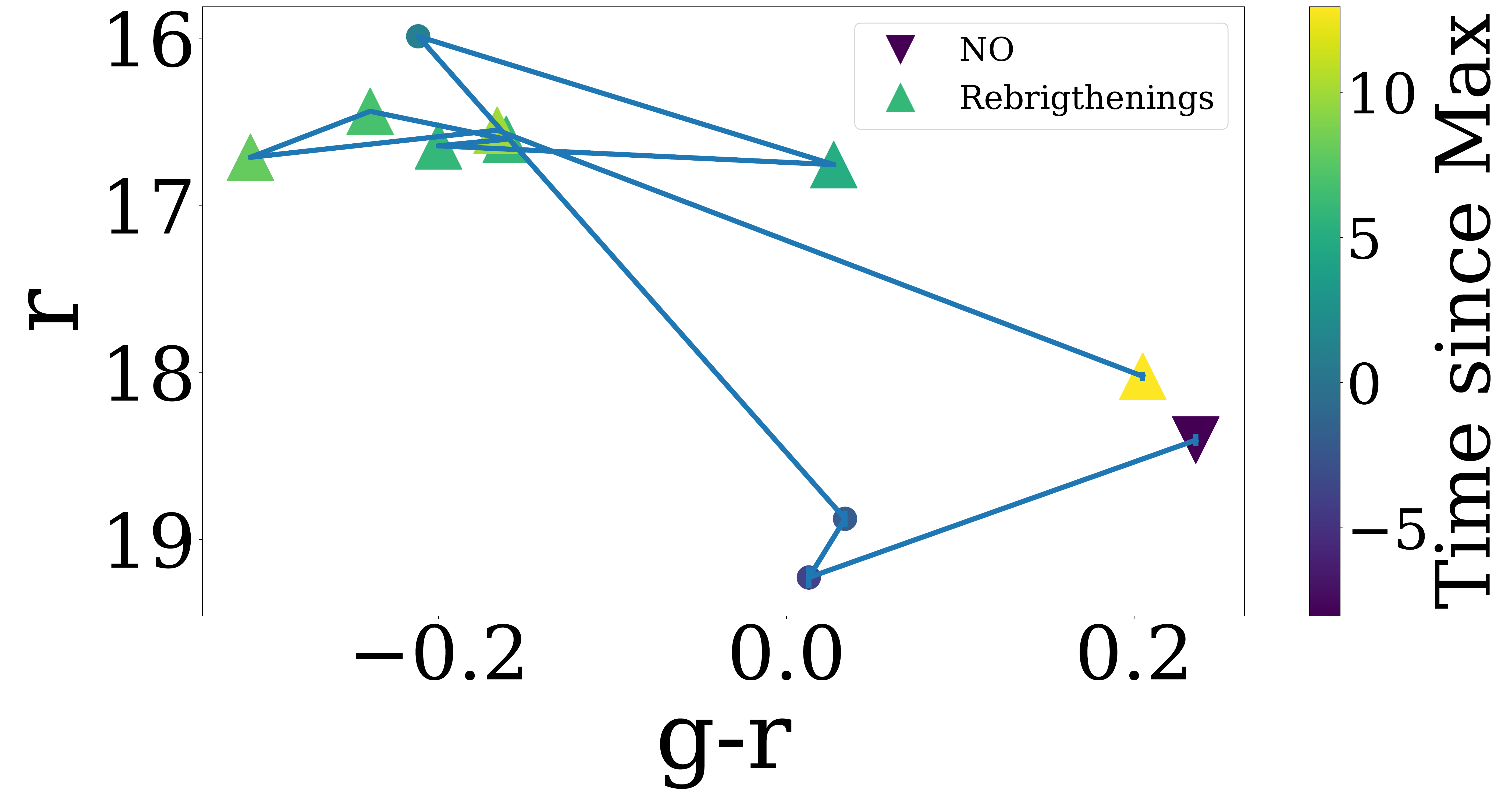}
    \caption{Evolution of the colour of PTF1 J0719+4858 from the ZTF points shown in fig.~\ref{fig:ptf0719ZTFandTESSshaded}. For each of the ZTF $r$ magnitudes, we select the closest $g$-band measurement in time, and require that they data points be taken $<$5 hours apart to be plotted. The inverted triangle corresponds to the normal outburst around 58846 MJD before the SO, the filled circles are the data points after the NO and until the end of the SO, and the upright triangles are the colours for the rebrightenings after the SO. The colours corresponds to the difference between the date of when the r value was taken and the maximum of the SO. Negative numbers indicate that data were obtained before the maximum of the SO, and positive numbers refer to dates after the maximum of the SO. The coverage of the SO is not complete but the data show that as the SO develops, the source is bluer and brighter, returning to redder colours at the end of the echo outbursts.  }
    \label{fig:ptf0719ZTFandTESScolour}
\end{figure}

This source was also observed simultaneously with ZTF in two different filters, $g$ and $r$,  (fig.~\ref{fig:ptf0719ZTFandTESSshaded}), allowing us to study the colour evolution of the SO as done in  \cite{RiveraSandoval60minOutburst2020arXiv201210356R}. Fig.~\ref{fig:ptf0719ZTFandTESScolour} shows the colours selected from the data points fig~.~\ref{fig:ptf0719ZTFandTESSshaded}. The colours of the outbursts are bluer than $g-r=0$, indicative of temperatures well above $10^4$~K.  

The first point in fig.~\ref{fig:ptf0719ZTFandTESScolour} is due to the very small outburst detected, which did not heat the disk enough to be the dominant component. However, as the SO occurs, the binary becomes bluer and brighter. The next point is redder compared to the plateau phase of the SO because it corresponds to data obtained during the first rebrightening. However, its colour indicates that the disk is still very hot. As the rest of the rebrightenings occur, the system becomes bluer and thus hotter. At the end of the series of rebrigtenings, the binary finally turns redder and fainter because it is going back to its quiescent colour. The behaviour of PTF1~J0719+4858 during SO is very similar to that of the 46 min AM~CVn SDSS~J141118+481257, but opposite to the one of the longer period AM~CVns SDSS~1137 and SDSS~0807 \citep{RiveraSandoval60minOutburst2020arXiv201210356R}, indicating that the triggering mechanism of the SO is very likely not the same that in these two long-period systems.

\subsection{KL Dra: a frequently outbursting AM~CVn shows further evidence for enhanced mass transfer}

The object KL~Dra was first discovered in outburst as a candidate supernova \citep{SchwartzSNKlDRa1998IAUC.6982....1S}.   Relatively soon after its discovery, however, it was identified via spectroscopy as a AM~CVn system \citep{Jha1998KLDraIAUC.6983....1J} and since then, it has been studied extensively in many wavelengths including X-ray and optical. \citep[e.g][]{WoodKlDraSumperHump2002MNRAS.334...87W,RamsayKLDra2010MNRAS.407.1819R}. It has an orbital period of 25 minutes, superoutburst recurrence time of about 60 days, and the typical superoutburst duration has been reported to be $\sim 15$ days by \cite{RamsayKLDra2010MNRAS.407.1819R} and more recently to be $10 \pm 0.7$ days by \cite{DuffyTESSKLDRA2021}, although we re-visit the issue of the outburst duration here.  All sectors for KL~Dra were recorded with 30 minutes cadence so we cannot perform searches for superhumps or orbital periodicity.  

The \emph{TESS} LC of this source was first studied by \cite{DuffyTESSKLDRA2021} where they reported for the first time evidence for precursors in AM~CVn systems. KL~Dra was continuously observed from sectors 14 to 26 (58682.87-59034.62 MJD). There were two sectors (15, 17) in which we could not resolve the source from the background.

\begin{figure}
	\includegraphics[width= 1.0 \columnwidth]{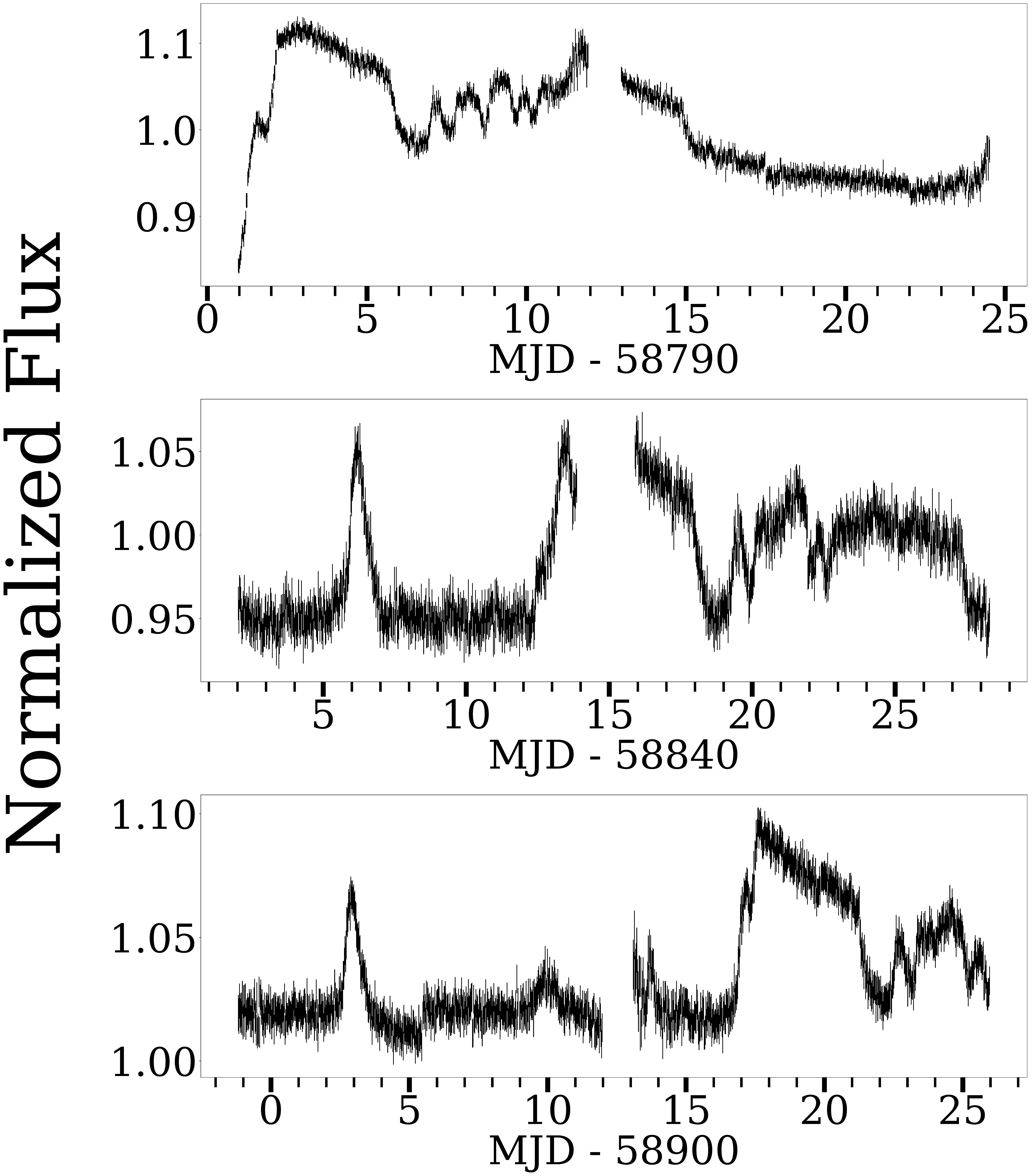}
    \caption{\emph{TESS} LC of three SO of KL Dra from sector 18 (top panel), 20 (middle panel) and 22 (bottom panel) at 30-minute cadence. All three SOs show similar form with a clear precursor before the SO, and then a series of rebrightenings lasting for $\sim 9$ days of variable length and brightness.}
    \label{fig:KLDPanel}
\end{figure}

We focused on the SO properties and compare them to other AM~CVns also observed by \emph{TESS}. Fig.~\ref{fig:KLDPanel} shows examples of SOs from KL~Dra observed by \emph{TESS}. Remarkable similarities are observed between these LCs and that of PTF~J0719+4858 (Fig.~\ref{fig:ptf0719ZTFandTESSshaded}), particularly in the shapes of the echo outbursts after the plateau phase of the SO. Similar to PTF~J0719+4858, the duration of the echo outbursts seems to be variable, with increasing duration and amplitudes for the later echoes relative to the first echoes (fig.~\ref{fig:KLDPanel}). In both systems a precursor is also observed. Given the similarities between KL~Dra and PTF~J0719+4858, one can apply analogous arguments to explain the behaviour of KL~Dra.  

We calculate the duration of the SOs. For sector 18 we get a duration of the SO of $\sim 7$ days from 58789.17 to 58796.18 MJD. This is substantially shorter than the value ($\sim 15$ days) reported previously for KL~Dra by \cite{RamsayKLDra2010MNRAS.407.1819R} and closer to the reported value by \citep{DuffyTESSKLDRA2021}. The sector 18 rebrightenings do, however, prolong the total period of activity by an additional 9.5 days to a total of about $\sim 16$ days including the SO. The outburst in sector 20  (see fig. \ref{fig:KLDPanel}) shows similar properties to that in sector 18 -- a 6 day superoutburst, extended to 15 days by rebrightenings -- and the full supercycle in sector 22 is only partially covered by \emph{TESS}, but it shows the main superoutburst to be similar to those in sectors 18 and 22.  
Most likely, then, the superoutbursts' durations have been overestimated due to the relatively poor cadence of the ground-based campaigns used to estimate the typical duration, making the echo outbursts to look as if they were part of the main SO. This highlights the need for observatories like \emph{TESS} that have continuous follow up of these outbursts, and also indicates that high cadence campaigns on the tails of superoutbursts with ground-based networks of telescopes like AAVSO for bright sources, or LCOGT for fainter objects, would be well-motivated.

\subsection{CP Eri: A short period AM~CVn with a short precursor and rebrigthenings}

CP Eri is an AM~CVn with an orbital period of 28 min \citep{Howell1991CPEri,Abbott1992CPEri}. It was in the field of view of \textit{TESS} in sectors 4 and 31. However, outburst activity was only detected in sector 4, for which the data were acquired with a 30-minute cadence from 58410 MJD (2018-10-19) to 58436 MJD (2018-11-14). 

\begin{figure}
	\includegraphics[width= 1.0 \columnwidth]{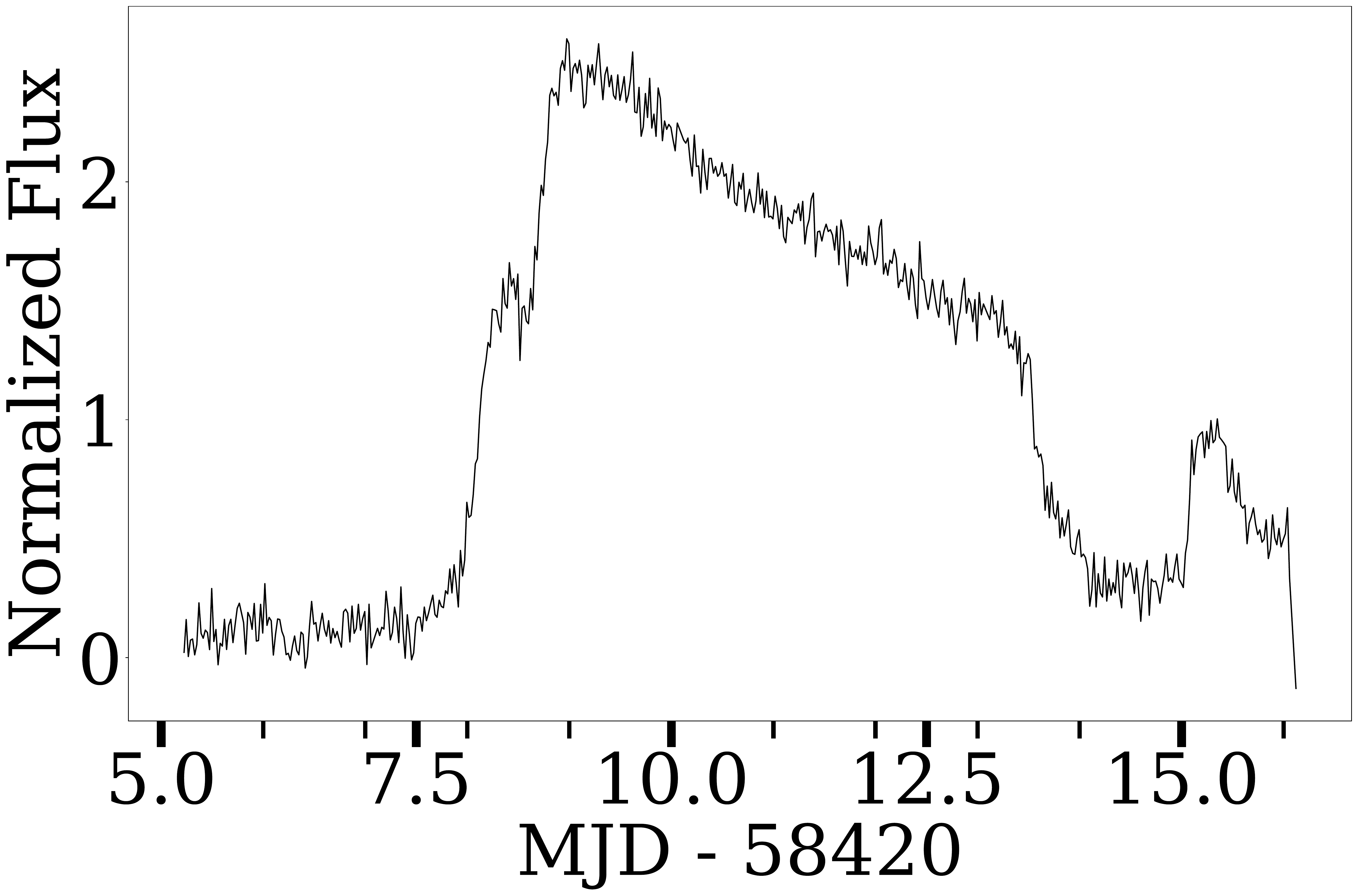}
    \caption{\emph{TESS} LC from sector 4 for CP Eri at 30-minute cadence. The LC shows a precursor, the SO and at least one echo outburst before the LC ends.}
    \label{fig:CPEriLCZTFandTESS}
\end{figure}

We detected a SO with a precursor in sector 4 (fig. \ref{fig:CPEriLCZTFandTESS}) which lasted for 5.8 days. The outburst duration is much shorter than the one previously reported by \citet{Levitan2015longterm}, which was $\sim 20$ days. The duration is also much shorter than the one expected from the classical DIM \citep{Cannizzo2019Duration}. A rebrightening was also detected shortly before the end of the data acquisition, but very likely more rebrightenings occurred which were not recorded. The SO and the rebrightening found with \emph{TESS} were also detected by \emph{ATLAS} \citep{Atlas2018PASP..130f4505T} and ZTF. The characteristics of this source are consistent with the pattern we are finding with the \emph{TESS} data in which the superoutburst durations seem to be previously overestimated using low-cadence ground-based data because the rebrightenings are merged together with the superoutburst. Due to the sampling frequency, a superhump analysis was not performed.




\subsection{SDSS J1043+5632: An AM CVn in SO with a series of short-lived echo outbursts}

SDSS~J1043+5632 was confirmed as an AM~CVn via spectroscopy by \cite{Carter2013MNRAS.429.2143C}. Based on the properties of its outbursts in the long-term study carried out by \cite{Levitan2015longterm}, an orbital period of 28.5 min was estimated for this source. However, a detailed study of the outbursts of this AM~CVn has not been performed until now. 

\emph{TESS} observed this binary in sector 21 from 2020-01-21 to 2020-02-18 at 30-minute cadence. A superoutburst with a precursor, together with a series of 8 echo outbursts were observed. The SO had a duration of 6.04 days, in agreement with the 55-day upper limit determined by \cite{Levitan2015longterm}. Unfortunately, the \emph{TESS} observations did not cover the entire series of echoes, likely leaving unobserved several of them. For the echoes that are visible in fig.~\ref{fig:SDSS1043ZTFandTESSShaded}, we note that they all have very similar (sharp) shapes, amplitudes and short duration (1.1 d). The similarity of that part of the light curve with the recently reported model by \cite{Hameuryrebright2021} for DNe of the type WZ~Sge is remarkable. These authors explained such behaviour by invoking the EMT model. The decay part of the SO is due to a decrease in the mass transfer rate, but in order to reproduce the series of echoes, the mass transfer from the donor needs to be constant for a period of several weeks in DNe (where the echoes occur), which in AM~CVns would be of a few days considering the smaller disks. The origin of that behaviour in the mass transfer rate is still unknown, but the fact that after the SO the binary remains above the original quiescent level strongly suggests changes in the mass transfer rate.

It is important to note that in the LC of SDSS~J1043+5632 there is no quiescent period between the several echoes. As occurs in the case of WZ~Sge \citep{Hameuryrebright2021}, that behaviour suggests that the cooling front that propagates from the outer part of the disk towards the inner one is reflected by a heating front, thus making the disk hot. Some of the factors that would influence such behaviour are a hot and small WD (and hence massive) and a lack of truncation in the disk, in order to keep a small inner disk radius \citep{Hameuryrebright2021,Dubus2001A&A...373..251D}.

\subsubsection{The colour evolution of SDSS~J1043+5632}

\begin{figure}
	\includegraphics[width= 1.0 \columnwidth]{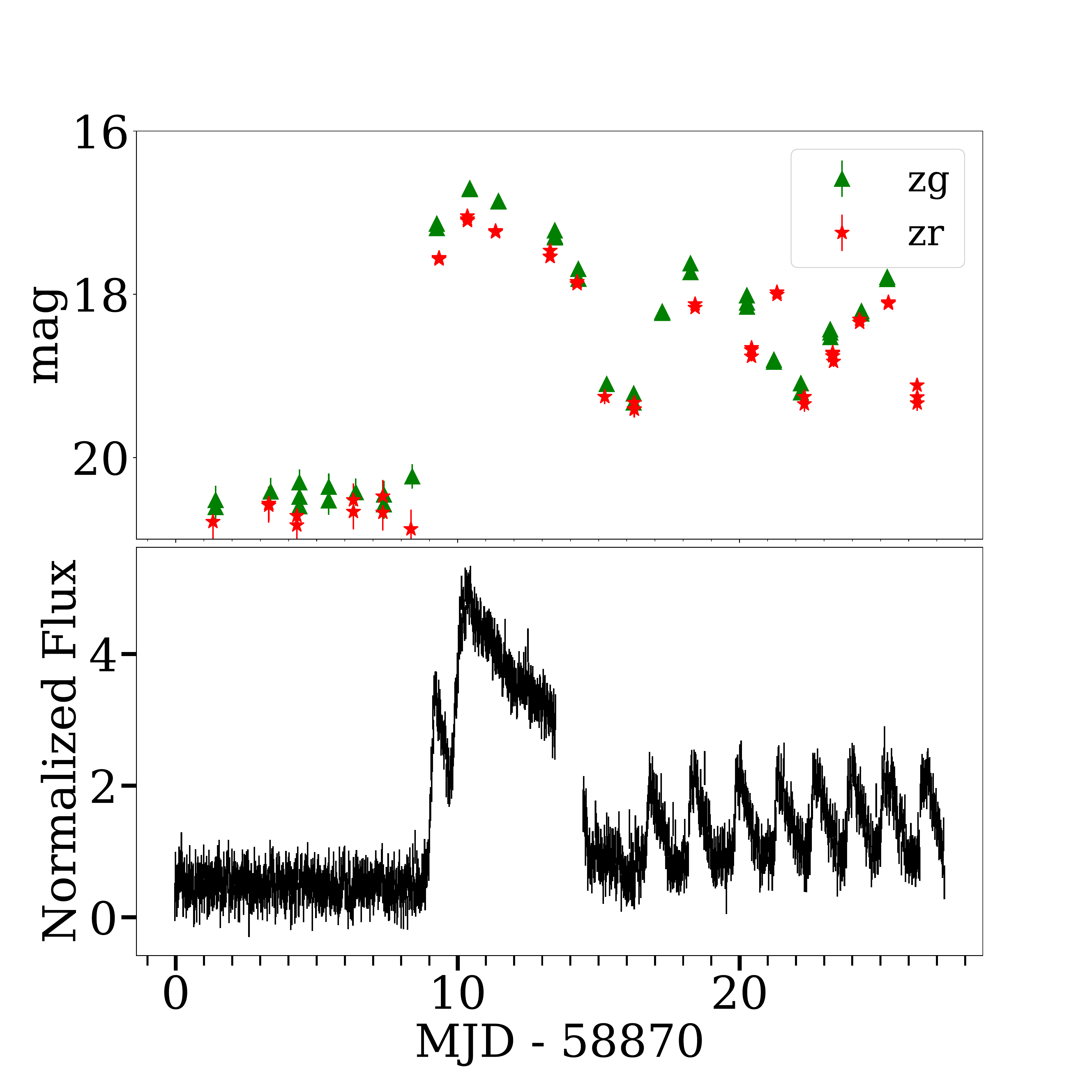}
    \caption{ZTF (top) and \emph{TESS} (bottom) LC at 30-minute cadence for SDSS J1043+5632 from sector 21. The \emph{TESS} LC shows the capability of \emph{TESS} and the month-long continuous coverage of the SO and the rebrigthenings. The \emph{TESS} LC shows a clear precursor, the SO and a series of short-lived ($\sim 1$ day) echo outbursts lasting for at least 10 days.  }
    \label{fig:SDSS1043ZTFandTESSShaded}
\end{figure}

\begin{figure}
	\includegraphics[width= 1.0 \columnwidth]{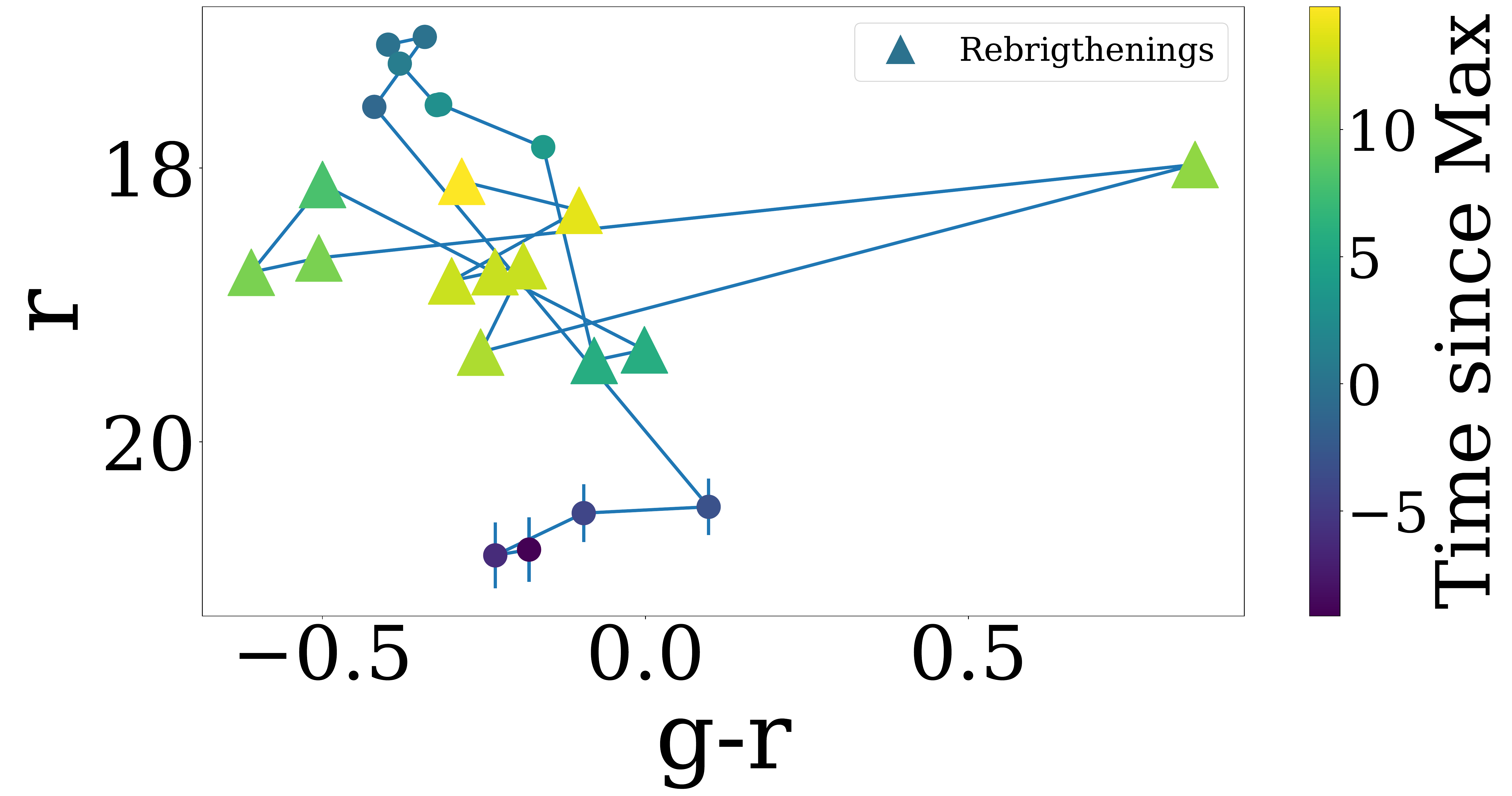}
    \caption{Evolution of the colour of SDSS~J1043+5632 from the ZTF points shown in fig.~\ref{fig:SDSS1043ZTFandTESSShaded}. The filled circles are the data points obtained before the SO and until its end. The triangles indicate the rebrightenings after the end of the SO. The colours corresponds to the difference between the date of the when the $r$ value was taken and the maximum of the SO. The colour evolution goes from redder and fainter (before the SO) to bluer and brighter (at the SO). During the rebrigthening phase then the system goes from a period of bluer but fainter to redder and fainter. The right most data point at $r = 17.9$ and $g-r = 0.8$ is probably not physical and better coverage of colours is needed for this source.  }
    \label{fig:SDSS1043coloursfromZTF}
\end{figure}

This source was also observed simultaneously with ZTF in two different filters, g and r,  (fig.~\ref{fig:SDSS1043ZTFandTESSShaded}), allowing us to track the colour evolution of the SO. Fig.~\ref{fig:SDSS1043coloursfromZTF} shows the colours selected from the data points of fig.~\ref{fig:SDSS1043ZTFandTESSShaded}. 
The initial colour behaviour of SDSS~J1043+5632 is similar to that described by \cite{2020HameuryDNe} for DNe, where the binary initially turns redder and brighter, because despite the fact that the accretion disk is becoming hotter,  the contribution of the hot spot is larger. Then, as the SO occurs the system becomes bluer and brighter. After the SO it becomes fainter and redder until the series of echo outbursts starts. The binary then turns again bluer and brighter. We note that the system is, remarkably, bluer than at similar luminosities during superoutbursts, perhaps indicating that the accretion disk spread to larger outer radius during the superoutbursts, and contracted again before the echoes set in. The end of the evolution of the echo outbursts was not depicted in fig.~\ref{fig:SDSS1043coloursfromZTF} due to the lack of coverage with \emph{TESS}. As in the case of PTF1~J0719+4858 the behaviour of SDSS~J1043+5632 is opposite to the one of the long period systems (SDSS~1137 and SDSS~0807), but more similar to the one of SDSS~J141118+481257 \citep{RiveraSandoval60minOutburst2020arXiv201210356R}. This provides further evidence that the colour behaviour is related to the mechanism that triggers the outburst.

\subsection{Gaia 16all: Possible 30-minute period AM CVn with two SOs, normal outburst and dozens of rebrigthenings}

This source was discovered in outburst by the Gaia Photometric Science Alerts \citep{GaiaAlert2012,Gaiaalert2013} on 2016-04-18 \citep{Delgado2016Gaia16all}. Gaia 16all has two sets of contiguous data. The sectors are contaminated with background stars, the clear periodicity present in the LC is due to a background W~UMa. The first set consists of sectors 1 through 13, from 58324.8 MJD (2018-07-25) to 58681.8 (2019-07-17), with a total duration of 357 days. During this period there were two superoutbursts detected, one in sector 5 and another one in sector 12. Their peaks are separated by 203 days, occurring on 58443.8 MJD (2018-11-21) and 58643.3 MJD (2019-06-09), respectively.

\begin{figure}
	\centering
    \textbf{Gaia 16all}\par\medskip
	\includegraphics[width= 1.0 \columnwidth]{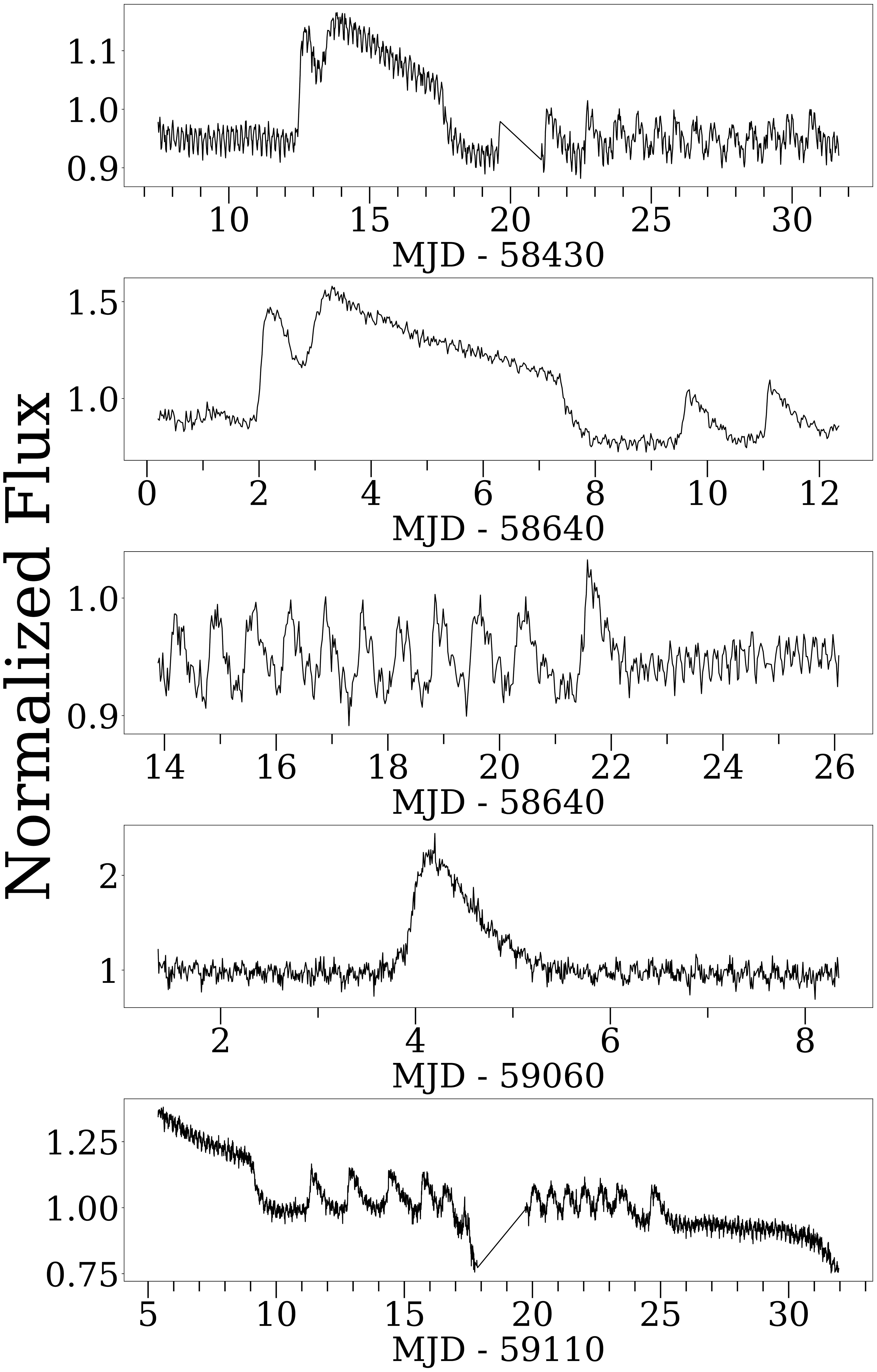}
    \caption{LC for sectors 5, 12, 13, 28 and 30 (top to bottom) from \emph{TESS } for Gaia 16all at 30-minutes cadence (Panel 1-3 top to bottom) and 2-minutes cadence (last 2 bottom panels). The visible variability in the LC is due to a W~UMa variable in the same pixel (21 arcsec) as the AM~CVn. The dip on sector 30 (bottom panel) is instrumental and due to the background. }
    \label{fig:gaia16allPanel}
\end{figure}

The second set is 191 days long, and is composed of sectors 27 to 33. \textit{TESS} observed this source from 59035.8 MJD (2020-07-05) to 59227.1 MJD (2021-01-13) at 10-minutes and 2-minutes cadence. The gap between sectors 13 and 27 is 354 days long and it is longer than the superoutburst recurrence time observed in the first data set. We see an outburst in sector 28 with a peak at 590674.2 MJD (2020-08-03; fig.~\ref{fig:gaia16allPanel}). This outburst was followed by an SO seen decaying at the beginning of sector 30 with the maximum at 59116.4 MJD (bottom panel of fig.~\ref{fig:gaia16allPanel}).

We use the 2-minute cadence data from sector 30 to search for periodicity. We first detrend the LC of that sector by fitting a polynomial to the SO plateau phase and perform Lomb-Scargle and Phase Dispersion Minimization. From the Lomb-Scargle we get a candidate period of $30.14 \pm 0.07$ min and a FAP of 0.017 using the bootstrap method. Using the PDM we get a similar period of $30.12$ min and a FAP of $2.5 \times 10^{-13}$ using the method described above in the timing analysis section. The folded LC at the period found with PDM is nonsinusoidal (fig.~\ref{fig:twomfoldedgaia16all}) and this could explain the higher significance found via the PDM method.

We can compare this period to the predicted period based on Gaia 16all outburst recurrence time. \cite{Levitan2015longterm} found the outburst recurrence time and orbital period are related by:

$$y = \alpha P^\beta_{orb} + \gamma$$

where $y$ is the outburst recurrence time, $P_{orb}$ is the orbital period in minutes, and $\alpha$, $\beta$, $\gamma$ are fit parameters. \cite{Levitan2015longterm} found the best-fitting values to be $\alpha = 1.53\times 10^{-9}$, $\beta=7.35$, and $\gamma$ = 24.7.  For a recurrence time of 203 days from the outbursts from sectors 5 and 13,  we get a predicted $P_{orb}$ to be 32 minutes, close to the peak of the periodogram found for the source.

Gaia 16all is thus found to be the system with the longest orbital period system with known normal outbursts and superoutbursts. This shows that AM~CVn with long recurrence time (in the orders of ~200 days) show normal outbursts as well as superoutbursts, suggesting that even at longer periods AM~CVns behave like SU~Uma as opposed to WZ~Sge and show both types of outbursts. 

The LC of Gaia 16all resembles the LC of SDSS J1043+5632. Both show a clear precursor before the SO and many echo outbursts after the end of the SO. For Gaia 16all this is visible in all three SOs where each SO is followed by a series of rebrightenings lasting from $\sim 12-15$ days in total and each individual rebrightening with a duration on the order of $\sim 1$ day.

The duration of the three SOs observed with \emph{TESS} are of 5.4, 5.5 and $>3$ days. These are about $\sim 1$ day shorter than the duration for SDSS~J1043+5632 that showed similar pattern. Neither period has been spectroscopically confirmed but SDSS~J1043+5632 is predicted to have shorter orbital period and shows longer outburst durations. This contradicts the relation found by \cite{Levitan2015longterm}, but given that the periods and durations are relatively similar, it may just be indicative of mild scatter in a relatively robust relation.

\begin{figure}
	\includegraphics[width= 1.0 \columnwidth]{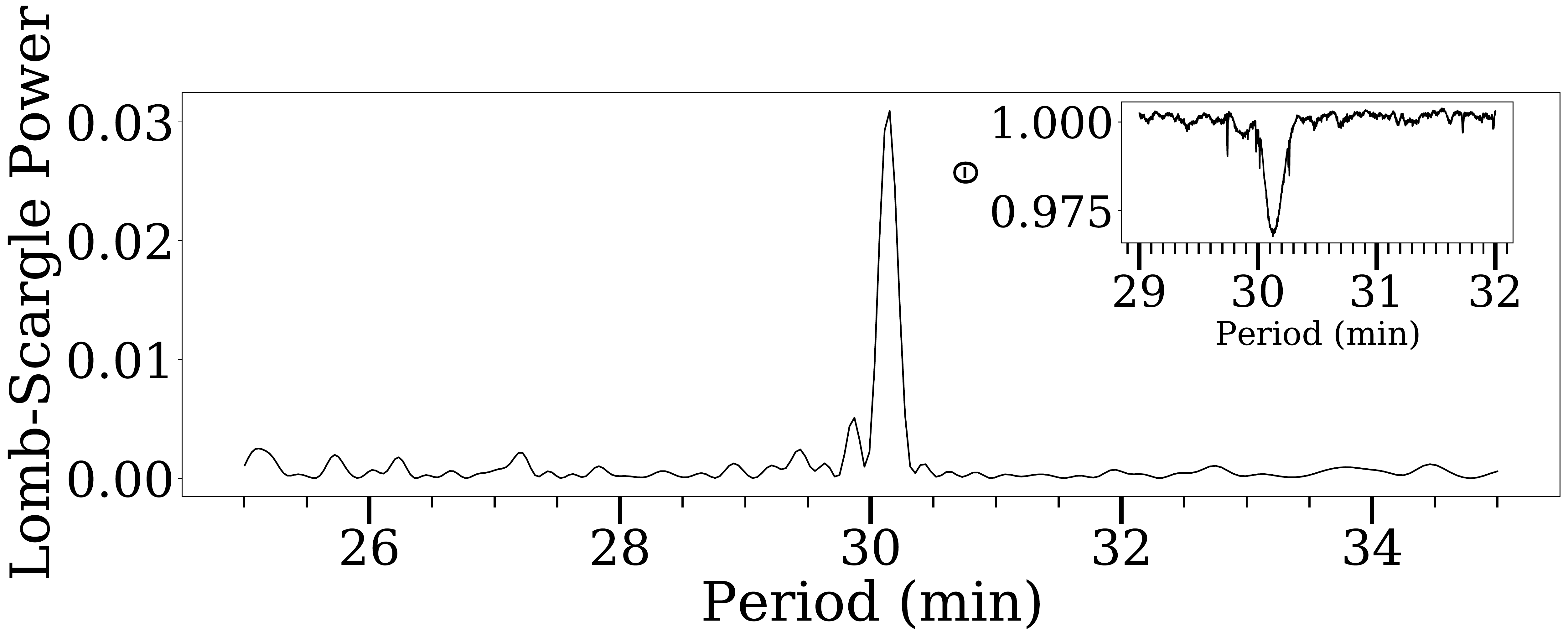}
    \caption{Lomb-Scargle for the 2-minute cadencde LC of  Gaia 16all. The periodogram clearly shows an isolated peak at 30.1 minutes. This period is also recovered using PDM (inset).}
    \label{fig:gaia16allperiodogramtwomin}
\end{figure}

\begin{figure}
	\includegraphics[width= 1.0 \columnwidth]{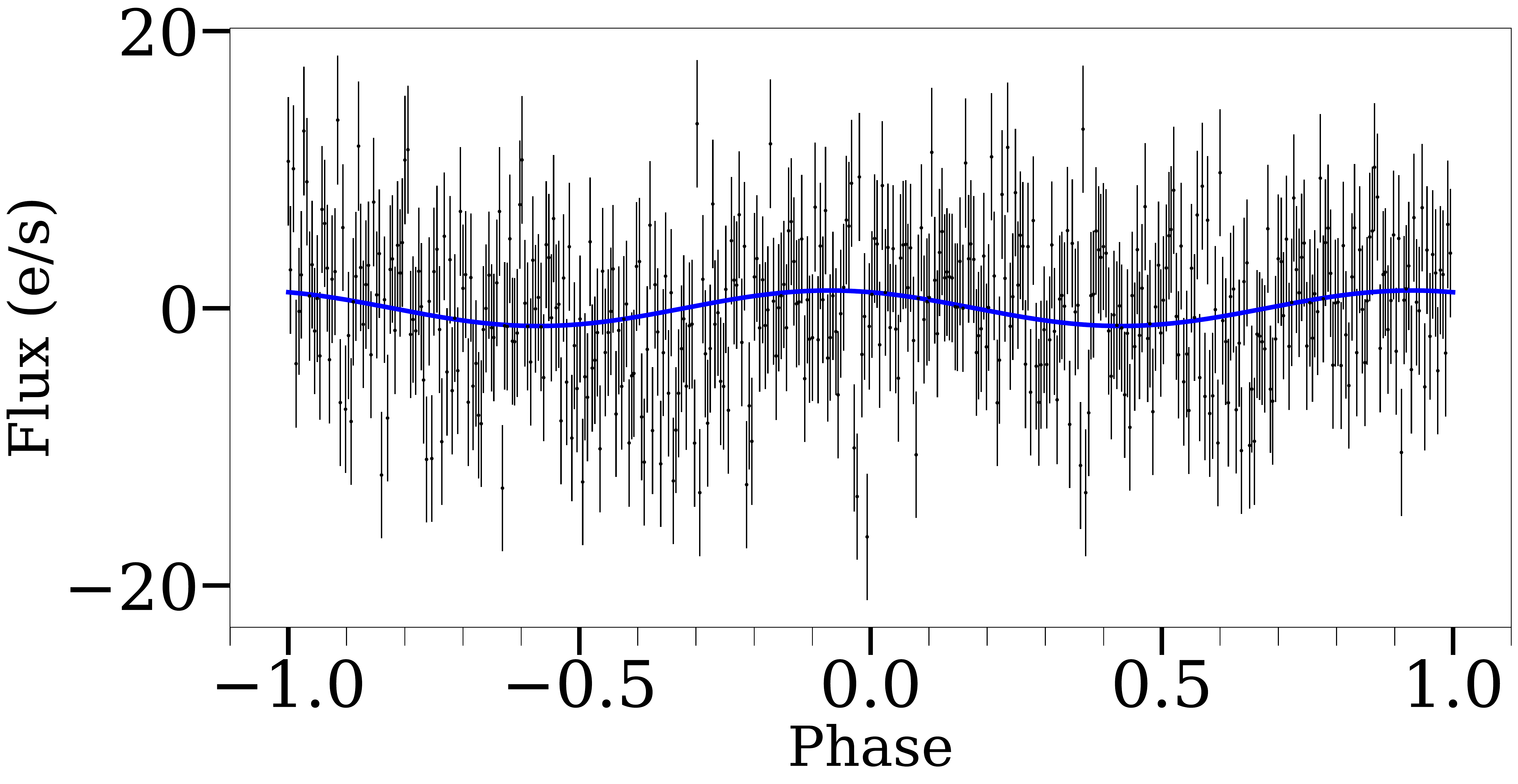}
    \caption{2-minutes cadence LC folded at 30.12 min for Gaia 16all. We plot only every 10th data point to avoid crowding. The LC is not strictly sinusoidal and could explain the higher significance found using the PDM method. The blue line corresponds to the best-fit sinusoid from the Lomb-Scargle. }
    \label{fig:twomfoldedgaia16all}
\end{figure}

\subsection{ZTF18abihypg}

\begin{figure}
\includegraphics[width= 1.0 \columnwidth]{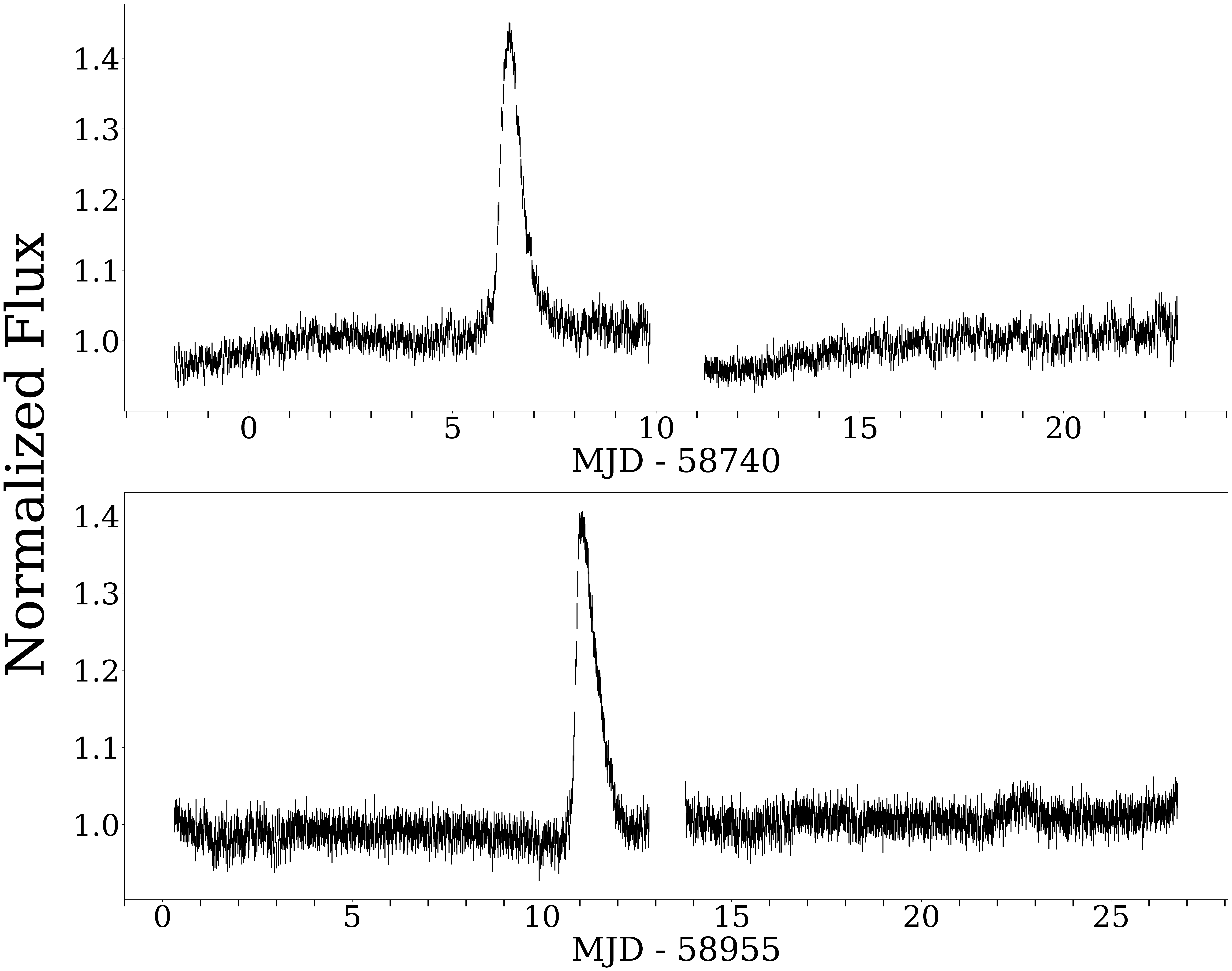}
    \caption{\emph{TESS} LC for the two outbursts detected by for ZTF18abihypg at 30-minute cadence. The top panel corresponds to sector 16 and the bottom one to sector 24.}
    \label{fig:panelZTF18abihypg}
\end{figure}

This source was found as an outbursting source in ZTF by \cite{Szkody2020AJ....159..198S}, and later confirmed via spectroscopy to be an AM~CVn by \cite{vanroestel2021}, who also looked at the outburts properties using ZTF data. Here we report the observations with \emph{TESS}. ZTF18abihypg was observed by \emph{TESS} in 3 sectors (16,17,24) showing two outbursts in sector 16 and sector 24 (fig.~\ref{fig:panelZTF18abihypg}). Both outbursts show similar profiles and duration of the order of one day. \cite{vanroestel2021} report for this source a high frequency of outbursts, but during the two contiguous sectors of observation by \emph{TESS} (~54 days) we only see one outburst, meaning that for this source the recurrence time for normal outburst is larger than 50 days. The cadence of both sectors was of 30-minutes, meaning that we could not perform any periodicity search on the outburst to look for possible superhumps or confirm the orbital period. Based on the SO recurrence time, \cite{vanroestel2021} predicts a period of $\sim 31.2$ minutes, this would imply a long recurrence time for the SO \citep{Levitan2015longterm}, and explains why we did not detect any SO in the two contiguous sectors.

\begin{figure}
	\includegraphics[width= 1.0 \columnwidth]{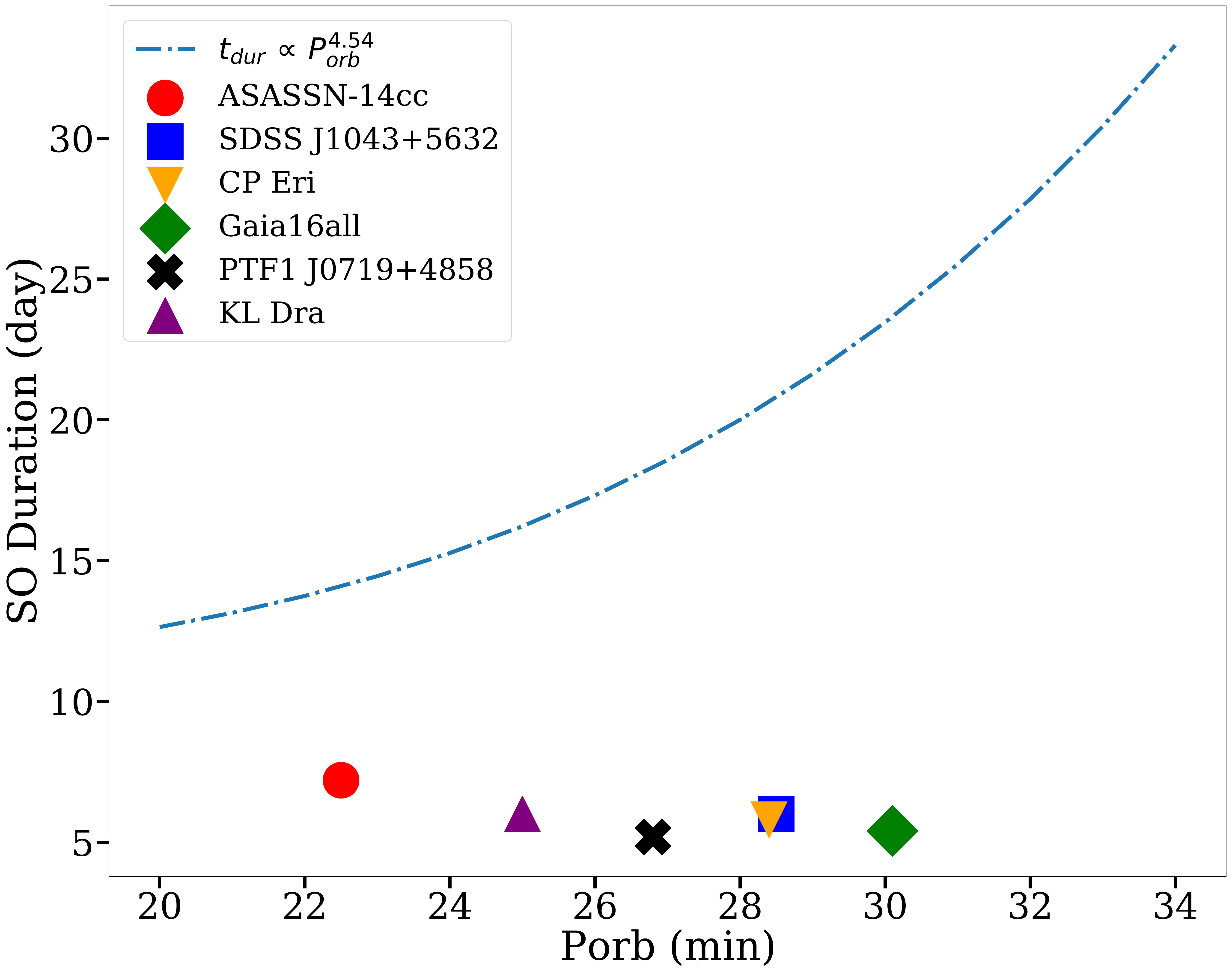}
    \caption{Superoutburst durations vs the orbital period for six sources with SO detected by \emph{TESS}. The blue dash-dotted line represents the relation found by \citet{Levitan2015longterm}
    . Due to the short time scales of the SO and the echo outbursts, and lack of good coverage as the one provided by \emph{TESS}, the duration of the SO  for many AM CVns might have been overestimated with scarce ground base data. Most likely, then, the superoutbursts' durations have been overestimated due to the relatively poor cadence of the ground-based campaigns used to estimate the typical duration, making the echo outbursts to look as if they were part of the main SO. Good coverage of the SO of systems with longer orbital period could provide good insight and constraints for models.  }
    \label{fig:plateaudurationvsporb}
\end{figure}

\subsection{Sources without detected superoutbursts}

Other AM CVns were observed by \emph{TESS}, but no outbursts or superoutbursts were detected. These are shown in Table \ref{nondetect} of the Appendix. Our sample includes several objects at short periods which are known to be persistently in bright states. Most of the systems have long orbital periods, such that even with many sectors observed, an outburst would have low probabilities to be detected considering the long recurrence times. The \citet[][]{Levitan2015longterm} relation between orbital period and superoutburst recurrence time finds that systems with periods longer than about 35 minutes have recurrence times longer than one year.  A few of the remaining objects have sufficient number of sectors of data to expect that outbursts would have taken place, but these systems are all sufficiently distant, and had previous outbursts sufficiently faint, that it is expected that they would not be detectable by \emph{TESS} because they were too faint.

\section{Summary}

\subsection{Precursors}

Precursors in SOs of AM~CVn were first reported for one system, KL~Dra, from \emph{TESS} data by  \citep{DuffyTESSKLDRA2021}. Here we extend the number of sources that show a clear precursor before the SO to 5. For the 6 systems we observed a SO, ASASSN-14cc was the only system without a clear precursor before the plateau phase of the SO. This might suggest that precursors are a common feature of SOs in AM~CVn systems. We also note that the only source without a precursor, ASASSn-14cc, is the system with the shortest orbital period in our sample. We emphasize that this is a small number of sources in the sample and of detected outbursts for each source, so further monitoring of AM~CVn systems and their NOs and SOs is needed to verify that in fact precursors are the norm in these types of systems, and to confirm if there is any possible dependence on the orbital period of AM~CVns.

\subsection{Echo Outbursts}

Echo outbursts have been detected before in AM~CVns \citep[e.g.][]{Green2020MNRAS.496.1243G}.   Here we used the continuous observations of \emph{TESS} to study in more detail the echo outbursts of these systems. Of the 6 systems reported here with a SO, 5 of them (PTF1J0719+4858, KL Dra, CP Eri, SDSSJ1043+5632 and Gaia 16all) showed at least one clear rebrightening after the end of the SO. The only system that did not show any echo outburst after its SO detected by \emph{TESS} was ASASSN-14cc. Like precursors, this could mean that echo outbursts or rebrightening phases are common in SOs of AM~CVns.
We see echo outbursts for systems of orbital period from 25-32 minutes, but systems with longer orbital period (43,46 minutes) also have shown echo outbursts discovered by other facilities \citep{Green2020MNRAS.496.1243G,2019RSM}, meaning that there is no apparent correlation between the orbital period and the appearance of echo outbursts.

While echo outbursts seem to be a common feature in SO of AM~CVns, we find that they show diverse behaviour. For two systems, KL Dra and PTF1J0719+4858, we find a sequence of rebrightenings of varying (and, in fact, increasing) duration from 0.7 days to about ~ 5 days that last for at least 9 days after the end of the SO. Most likely, then, the superoutbursts’ durations in prior work have been overestimated due to the relatively poor cadence of the ground-based campaigns used to estimate the typical durations, merging the echo outbursts with the main SO (Fig.\ref{fig:plateaudurationvsporb}). As discussed in the main text for both sources, the shape of the echo outbursts for these two systems points towards evidence for enhanced mass transfer in these binaries.

The other three systems (CP~Eri, SDSS~J1043+5632 and Gaia~16all) that showed echo outbursts showed different characteristics, they are more similar to the ones in the dwarf novae EG~Cancri and WZ~Sge \citep{Patterson1998EG_Candri,Patterson2002WZSGe}. They show a series of similar short-lived ($\sim1$ day) echo outbursts for up to 15 days after the SO. This could be explained due to an increase in the mass transfer rate in the DIM model \citep{Hameuryrebright2021}.

We stress again that this is still a small number of observed AM~CVns and SOs for each system. Therefore, additional monitoring is necessary to understand the different types of echo outburst in AM~CVns and characterize which parameters determine their nature, e.g. orbital periods, whether the donor is a white dwarf or a helium star, chemical composition, long term history of mass transfer and perhaps other parameters might all be important.

\subsection{Outbursts Colors}

For two sources, SDSS~J1043+5632 and PTF1~J0719+4858, we had simultaneous \emph{TESS} and ZTF data for at least part of the SO. Following the approaches by \cite{RiveraSandoval60minOutburst2020arXiv201210356R}, we look at the color evolution of these systems. Both systems showed similar behaviour to SDSS~J141118+481257 in that they become bluer as it moves on the rising phase of the outburst, and it is bluest and brightest during the peak. After the SO, the binary becomes redder and fainter reaching colours similar to those before the superoutburst. This behaviour is similar to dwarf novae outbursts and predicted by the DIM \citep{2020HameuryDNe}.

\section{Conclusions}

The sample size of sources and outbursts is still not sufficient to make broad statistical statements about the outburst properties, but even from this modest sample of continuous light curves, we can establish several new points of phenomenology of the transient AM~CVns.

\begin{enumerate}
    \item It is clear that there is a wide range of phenomenology in the outburst behaviour of AM~CVn systems. 
    
    \item Precursors seem to be more common than previously believed. Our data shows that given their extremely short duration these are easy to miss with ground-based imaging studies, explaining why the first precursor was only recently detected using \emph{TESS}.  This shows a previously unappreciated similarity to what had been found for normal CVs with {\it Kepler} \citep{Osaki2013KeplerV15}.

    \item  Normal outbursts are relatively common in many AM~CVns, and have previously been missed because their durations are often less than a full day, meaning that past light curve sampling was insufficient for robustly discovering them.
    \item Many systems show large numbers of echo outbursts associated with a single superoutburst, and frequently, the durations and separations of these increase with time which seems to be associated to variations in the mass transfer rate. 
    \item Superoutburst durations for many systems in the literature have likely been overestimated in poorly sampled light curves (fig.~\ref{fig:plateaudurationvsporb}), and represent the sum of the superoutburst duration, plus the duration of the time interval over which the echo outbursts were taking place.  This issue is more important for empirical tests of outbursts mechanism models than for understanding the detectability of the superoutbursts, but affects both.
    \item Candidate periods are found for two systems which previously did not have orbital periods, PTF1~J2219+135, which shows a marginally significant period of about 28 minutes, and Gaia 16all, which shows a much more significant period of 30.1 minutes.  In both cases, these periods are similar to what is expected based on outburst recurrence times, but spectroscopic confirmation would be beneficial.
    \item The colour evolution of the systems for which there was ZTF simultaneous coverage with \emph{TESS} shows that the objects become bluer and brighter when they are closer to the maximum of the superoutburst. That colour trend pattern is opposite the one followed by long-period AM CVns in which the triggering outburst mechanism seems to be EMT. This shows that despite the fact that EMT might be present at orbital periods shorter than 30 min, it does not seem to be the dominant triggering process. From that complementary ZTF data, it is clear that further simultaneous multiwavelength observations of AM~CVns in outbursts are needed to characterize the mechanisms involved through their colour evolution. 
    
\end{enumerate}

\section*{Acknowledgements}

We thank J.M. Hameury for helpful discussions and a careful reading of this manuscript. We thank the anonymous referee for a constructive report which has improved the clarity and quality of this paper. LERS was supported by an Avadh Bhatia Fellowship at the University of Alberta during much of the time period over which this work was done. This paper includes data collected with the \emph{TESS} mission, obtained from the MAST data archive at the Space Telescope Science Institute (STScI). Funding for the \emph{TESS} mission is provided by the NASA Explorer Program. STScI is operated by the Association of Universities for Research in Astronomy, Inc., under NASA contract NAS 5–26555. Based on observations obtained with the Samuel Oschin 48-inch Telescope at the Palomar Observatory as part of the Zwicky Transient Facility project. ZTF is supported by the National Science Foundation under Grant No. AST-1440341 and a collaboration including Caltech, IPAC, the Weizmann Institute for Science, the Oskar Klein Center at Stockholm University, the University of Maryland, the University of Washington, Deutsches Elektronen-Synchrotron and Humboldt University, Los Alamos National Laboratories, the TANGO Consortium of Taiwan, the University of Wisconsin at Milwaukee, and Lawrence Berkeley National Laboratories. Operations are conducted by COO, IPAC, and UW. \emph{Software used:} PyAstronomy \citep{pyastronomypackage}, SciPy \citep{SciPy-NMeth}, Astropy \citep{astropy:2013,astropy:2018}, Matplotlib \citep{Hunter:2007}, specutils \citep{specutilnicholas_earl_2021_4603801}.

\section*{Data Availability}

All data used in this document is public and it can be found in the ZTF and \emph{TESS} webpages.



\bibliographystyle{mnras}
\bibliography{amcvn} 

\begin{thebibliography}{}
\makeatletter
\relax
\def\mn@urlcharsother{\let\do\@makeother \do\$\do\&\do\#\do\^\do\_\do\%\do\~}
\def\mn@doi{\begingroup\mn@urlcharsother \@ifnextchar [ {\mn@doi@}
  {\mn@doi@[]}}
\def\mn@doi@[#1]#2{\def\@tempa{#1}\ifx\@tempa\@empty \href
  {http://dx.doi.org/#2} {doi:#2}\else \href {http://dx.doi.org/#2} {#1}\fi
  \endgroup}
\def\mn@eprint#1#2{\mn@eprint@#1:#2::\@nil}
\def\mn@eprint@arXiv#1{\href {http://arxiv.org/abs/#1} {{\tt arXiv:#1}}}
\def\mn@eprint@dblp#1{\href {http://dblp.uni-trier.de/rec/bibtex/#1.xml}
  {dblp:#1}}
\def\mn@eprint@#1:#2:#3:#4\@nil{\def\@tempa {#1}\def\@tempb {#2}\def\@tempc
  {#3}\ifx \@tempc \@empty \let \@tempc \@tempb \let \@tempb \@tempa \fi \ifx
  \@tempb \@empty \def\@tempb {arXiv}\fi \@ifundefined
  {mn@eprint@\@tempb}{\@tempb:\@tempc}{\expandafter \expandafter \csname
  mn@eprint@\@tempb\endcsname \expandafter{\@tempc}}}

\bibitem[\protect\citeauthoryear{{Abbott}, {Robinson}, {Hill}  \&
  {Haswell}}{{Abbott} et~al.}{1992}]{Abbott1992CPEri}
{Abbott} T. M.~C.,  {Robinson} E.~L.,  {Hill} G.~J.,   {Haswell} C.~A.,  1992,
  \mn@doi [\apj] {10.1086/171960}, \href
  {https://ui.adsabs.harvard.edu/abs/1992ApJ...399..680A} {399, 680}

\bibitem[\protect\citeauthoryear{{Antipova}}{{Antipova}}{1987}]{KPRelationAntipova1987Ap&SS.131..453A}
{Antipova} L.~I.,  1987, \mn@doi [\apss] {10.1007/BF00668125}, \href
  {https://ui.adsabs.harvard.edu/abs/1987Ap&SS.131..453A} {131, 453}

\bibitem[\protect\citeauthoryear{{Astropy Collaboration} et~al.,}{{Astropy
  Collaboration} et~al.}{2013}]{astropy:2013}
{Astropy Collaboration} et~al., 2013, \mn@doi [\aap]
  {10.1051/0004-6361/201322068}, \href
  {http://adsabs.harvard.edu/abs/2013A%26A...558A..33A} {558, A33}

\bibitem[\protect\citeauthoryear{{Astropy Collaboration} et~al.,}{{Astropy
  Collaboration} et~al.}{2018}]{astropy:2018}
{Astropy Collaboration} et~al., 2018, \mn@doi [\aj] {10.3847/1538-3881/aabc4f},
  \href {https://ui.adsabs.harvard.edu/abs/2018AJ....156..123A} {156, 123}

\bibitem[\protect\citeauthoryear{{Balbus} \& {Hawley}}{{Balbus} \&
  {Hawley}}{1991}]{BalbusHawley}
{Balbus} S.~A.,  {Hawley} J.~F.,  1991, \mn@doi [\apj] {10.1086/170270}, \href
  {https://ui.adsabs.harvard.edu/abs/1991ApJ...376..214B} {376, 214}

\bibitem[\protect\citeauthoryear{{Buat-M{\'e}nard} \&
  {Hameury}}{{Buat-M{\'e}nard} \&
  {Hameury}}{2002}]{BuatEMTmodel2002A&A...386..891B}
{Buat-M{\'e}nard} V.,  {Hameury} J.~M.,  2002, \mn@doi [\aap]
  {10.1051/0004-6361:20020307}, \href
  {https://ui.adsabs.harvard.edu/abs/2002A&A...386..891B} {386, 891}

\bibitem[\protect\citeauthoryear{{Burdge} et~al.,}{{Burdge}
  et~al.}{2020}]{Burdge2020ApJ}
{Burdge} K.~B.,  et~al., 2020, \mn@doi [\apj] {10.3847/1538-4357/abc261}, \href
  {https://ui.adsabs.harvard.edu/abs/2020ApJ...905...32B} {905, 32}

\bibitem[\protect\citeauthoryear{{Cannizzo}}{{Cannizzo}}{1984}]{Cannizzo1984NaturDIM}
{Cannizzo} J.~K.,  1984, \mn@doi [\nat] {10.1038/311443a0}, \href
  {https://ui.adsabs.harvard.edu/abs/1984Natur.311..443C} {311, 443}

\bibitem[\protect\citeauthoryear{{Cannizzo} \& {Nelemans}}{{Cannizzo} \&
  {Nelemans}}{2015}]{CannizzoNelemansDIM2015}
{Cannizzo} J.~K.,  {Nelemans} G.,  2015, \mn@doi [\apj]
  {10.1088/0004-637X/803/1/19}, \href
  {https://ui.adsabs.harvard.edu/abs/2015ApJ...803...19C} {803, 19}

\bibitem[\protect\citeauthoryear{{Cannizzo} \& {Ramsay}}{{Cannizzo} \&
  {Ramsay}}{2019}]{Cannizzo2019Duration}
{Cannizzo} J.~K.,  {Ramsay} G.,  2019, \mn@doi [\aj]
  {10.3847/1538-3881/ab04ac}, \href
  {https://ui.adsabs.harvard.edu/abs/2019AJ....157..130C} {157, 130}

\bibitem[\protect\citeauthoryear{{Cannizzo}, {Ghosh}  \& {Wheeler}}{{Cannizzo}
  et~al.}{1982}]{Cannizzo1982ApJ...260L..83C}
{Cannizzo} J.~K.,  {Ghosh} P.,   {Wheeler} J.~C.,  1982, \mn@doi [\apjl]
  {10.1086/183875}, \href
  {https://ui.adsabs.harvard.edu/abs/1982ApJ...260L..83C} {260, L83}

\bibitem[\protect\citeauthoryear{{Cannizzo}, {Smale}, {Wood}, {Still}  \&
  {Howell}}{{Cannizzo} et~al.}{2012}]{Cannizzo2012ApJPrecursor}
{Cannizzo} J.~K.,  {Smale} A.~P.,  {Wood} M.~A.,  {Still} M.~D.,   {Howell}
  S.~B.,  2012, \mn@doi [\apj] {10.1088/0004-637X/747/2/117}, \href
  {https://ui.adsabs.harvard.edu/abs/2012ApJ...747..117C} {747, 117}

\bibitem[\protect\citeauthoryear{{Carter} et~al.,}{{Carter}
  et~al.}{2013}]{Carter2013MNRAS.429.2143C}
{Carter} P.~J.,  et~al., 2013, \mn@doi [\mnras] {10.1093/mnras/sts485}, \href
  {https://ui.adsabs.harvard.edu/abs/2013MNRAS.429.2143C} {429, 2143}

\bibitem[\protect\citeauthoryear{{Cumming}}{{Cumming}}{2004}]{Cumming2004}
{Cumming} A.,  2004, \mn@doi [\mnras] {10.1111/j.1365-2966.2004.08275.x}, \href
  {https://ui.adsabs.harvard.edu/abs/2004MNRAS.354.1165C} {354, 1165}

\bibitem[\protect\citeauthoryear{{Czesla}, {Schr{\"o}ter}, {Schneider},
  {Huber}, {Pfeifer}, {Andreasen}  \& {Zechmeister}}{{Czesla}
  et~al.}{2019}]{pyastronomypackage}
{Czesla} S.,  {Schr{\"o}ter} S.,  {Schneider} C.~P.,  {Huber} K.~F.,  {Pfeifer}
  F.,  {Andreasen} D.~T.,   {Zechmeister} M.,  2019, {PyA: Python
  astronomy-related packages} (\mn@eprint {ascl} {1906.010})

\bibitem[\protect\citeauthoryear{{Delgado}, {Harrison}, {Hodgkin}, {Leeuwen},
  {Rixon}  \& {Yoldas}}{{Delgado} et~al.}{2016}]{Delgado2016Gaia16all}
{Delgado} A.,  {Harrison} D.,  {Hodgkin} S.,  {Leeuwen} M.~V.,  {Rixon} G.,
  {Yoldas} A.,  2016, Transient Name Server Discovery Report, \href
  {https://ui.adsabs.harvard.edu/abs/2016TNSTR.481....1D} {2016-481, 1}

\bibitem[\protect\citeauthoryear{{Dubus}, {Hameury}  \& {Lasota}}{{Dubus}
  et~al.}{2001}]{Dubus2001A&A...373..251D}
{Dubus} G.,  {Hameury} J.~M.,   {Lasota} J.~P.,  2001, \mn@doi [\aap]
  {10.1051/0004-6361:20010632}, \href
  {https://ui.adsabs.harvard.edu/abs/2001A&A...373..251D} {373, 251}

\bibitem[\protect\citeauthoryear{{Duffy} et~al.,}{{Duffy}
  et~al.}{2021}]{DuffyTESSKLDRA2021}
{Duffy} C.,  et~al., 2021, \mn@doi [\mnras] {10.1093/mnras/stab389}, \href
  {https://ui.adsabs.harvard.edu/abs/2021MNRAS.tmp..425D} {}

\bibitem[\protect\citeauthoryear{Earl et~al.,}{Earl
  et~al.}{2021}]{specutilnicholas_earl_2021_4603801}
Earl N.,  et~al., 2021, astropy/specutils: v1.2,
  \mn@doi{10.5281/zenodo.4603801}, \url
  {https://doi.org/10.5281/zenodo.4603801}

\bibitem[\protect\citeauthoryear{{El-Khoury} \& {Wickramasinghe}}{{El-Khoury}
  \& {Wickramasinghe}}{2000}]{ElKhoury2000A&A}
{El-Khoury} W.,  {Wickramasinghe} D.,  2000, \aap, \href
  {https://ui.adsabs.harvard.edu/abs/2000A&A...358..154E} {358, 154}

\bibitem[\protect\citeauthoryear{{Feinstein} et~al.,}{{Feinstein}
  et~al.}{2019}]{EleanorFeinstein2019}
{Feinstein} A.~D.,  et~al., 2019, \mn@doi [\pasp] {10.1088/1538-3873/ab291c},
  \href {https://ui.adsabs.harvard.edu/abs/2019PASP..131i4502F} {131, 094502}

\bibitem[\protect\citeauthoryear{{Fontaine} et~al.,}{{Fontaine}
  et~al.}{2011}]{Fontaine2011SDSS1908}
{Fontaine} G.,  et~al., 2011, \mn@doi [\apj] {10.1088/0004-637X/726/2/92},
  \href {https://ui.adsabs.harvard.edu/abs/2011ApJ...726...92F} {726, 92}

\bibitem[\protect\citeauthoryear{{Green} et~al.,}{{Green}
  et~al.}{2018}]{Green2018SDSS1351}
{Green} M.~J.,  et~al., 2018, \mn@doi [\mnras] {10.1093/mnras/sty1032}, \href
  {https://ui.adsabs.harvard.edu/abs/2018MNRAS.477.5646G} {477, 5646}

\bibitem[\protect\citeauthoryear{{Green} et~al.,}{{Green}
  et~al.}{2020}]{Green2020MNRAS.496.1243G}
{Green} M.~J.,  et~al., 2020, \mn@doi [\mnras] {10.1093/mnras/staa1509}, \href
  {https://ui.adsabs.harvard.edu/abs/2020MNRAS.496.1243G} {496, 1243}

\bibitem[\protect\citeauthoryear{{Hameury}}{{Hameury}}{2020}]{HameuryDIMReview2020AdSpR}
{Hameury} J.~M.,  2020, \mn@doi [Advances in Space Research]
  {10.1016/j.asr.2019.10.022}, \href
  {https://ui.adsabs.harvard.edu/abs/2020AdSpR..66.1004H} {66, 1004}

\bibitem[\protect\citeauthoryear{{Hameury} \& {Lasota}}{{Hameury} \&
  {Lasota}}{2021}]{Hameuryrebright2021}
{Hameury} J.~M.,  {Lasota} J.~P.,  2021, arXiv e-prints, \href
  {https://ui.adsabs.harvard.edu/abs/2021arXiv210402952H} {p. arXiv:2104.02952}

\bibitem[\protect\citeauthoryear{{Hameury}, {Knigge}, {Lasota}, {Hambsch}  \&
  {James}}{{Hameury} et~al.}{2020}]{2020HameuryDNe}
{Hameury} J.~M.,  {Knigge} C.,  {Lasota} J.~P.,  {Hambsch} F.~J.,   {James} R.,
   2020, \mn@doi [\aap] {10.1051/0004-6361/202037631}, \href
  {https://ui.adsabs.harvard.edu/abs/2020A&A...636A...1H} {636, A1}

\bibitem[\protect\citeauthoryear{{Han}, {Soonthornthum}, {Qian},
  {Sarotsakulchai}, {Zhu}, {Dong}  \& {Zhi}}{{Han}
  et~al.}{2021}]{Han2021NewA...8701604H}
{Han} Z.,  {Soonthornthum} B.,  {Qian} S.,  {Sarotsakulchai} T.,  {Zhu} L.,
  {Dong} A.,   {Zhi} Q.,  2021, \mn@doi [\na] {10.1016/j.newast.2021.101604},
  \href {https://ui.adsabs.harvard.edu/abs/2021NewA...8701604H} {87, 101604}

\bibitem[\protect\citeauthoryear{{Hodgkin}, {Wyrzykowski}, {Blagorodnova}  \&
  {Koposov}}{{Hodgkin} et~al.}{2013}]{Gaiaalert2013}
{Hodgkin} S.~T.,  {Wyrzykowski} L.,  {Blagorodnova} N.,   {Koposov} S.,  2013,
  \mn@doi [Philosophical Transactions of the Royal Society of London Series A]
  {10.1098/rsta.2012.0239}, \href
  {https://ui.adsabs.harvard.edu/abs/2013RSPTA.37120239H} {371, 20120239}

\bibitem[\protect\citeauthoryear{{Horne} \& {Baliunas}}{{Horne} \&
  {Baliunas}}{1986}]{HorneLombFreq1986}
{Horne} J.~H.,  {Baliunas} S.~L.,  1986, \mn@doi [\apj] {10.1086/164037}, \href
  {https://ui.adsabs.harvard.edu/abs/1986ApJ...302..757H} {302, 757}

\bibitem[\protect\citeauthoryear{{Howell}, {Szkody}, {Kreidl}  \&
  {Dobrzycka}}{{Howell} et~al.}{1991}]{Howell1991CPEri}
{Howell} S.~B.,  {Szkody} P.,  {Kreidl} T.~J.,   {Dobrzycka} D.,  1991, \mn@doi
  [\pasp] {10.1086/132819}, \href
  {https://ui.adsabs.harvard.edu/abs/1991PASP..103..300H} {103, 300}

\bibitem[\protect\citeauthoryear{Hunter}{Hunter}{2007}]{Hunter:2007}
Hunter J.~D.,  2007, \mn@doi [Computing in Science \& Engineering]
  {10.1109/MCSE.2007.55}, 9, 90

\bibitem[\protect\citeauthoryear{{Imada}, {Izumiura}, {Kuroda}, {Yanagisawa},
  {Kawai}, {Omodaka}  \& {Miyanoshita}}{{Imada}
  et~al.}{2012}]{Imada2012PASJ...64L...5I}
{Imada} A.,  {Izumiura} H.,  {Kuroda} D.,  {Yanagisawa} K.,  {Kawai} N.,
  {Omodaka} T.,   {Miyanoshita} R.,  2012, \mn@doi [\pasj]
  {10.1093/pasj/64.5.L5}, \href
  {https://ui.adsabs.harvard.edu/abs/2012PASJ...64L...5I} {64, L5}

\bibitem[\protect\citeauthoryear{{Jenkins} et~al.,}{{Jenkins}
  et~al.}{2016}]{TESSSPOC2016SPIE}
{Jenkins} J.~M.,  et~al., 2016, in {Chiozzi} G.,  {Guzman} J.~C.,  eds,
  Society of Photo-Optical Instrumentation Engineers (SPIE) Conference Series
  Vol. 9913, Software and Cyberinfrastructure for Astronomy IV. p. 99133E,
  \mn@doi{10.1117/12.2233418}

\bibitem[\protect\citeauthoryear{{Jha}, {Garnavich}, {Challis}, {Kirshner}  \&
  {Berlind}}{{Jha} et~al.}{1998}]{Jha1998KLDraIAUC.6983....1J}
{Jha} S.,  {Garnavich} P.,  {Challis} P.,  {Kirshner} R.,   {Berlind} P.,
  1998, \iaucirc, \href {https://ui.adsabs.harvard.edu/abs/1998IAUC.6983....1J}
  {6983, 1}

\bibitem[\protect\citeauthoryear{{Kato} \& {Osaki}}{{Kato} \&
  {Osaki}}{2013}]{2013PASJ...65...97K}
{Kato} T.,  {Osaki} Y.,  2013, \mn@doi [\pasj] {10.1093/pasj/65.5.97}, \href
  {https://ui.adsabs.harvard.edu/abs/2013PASJ...65...97K} {65, 97}

\bibitem[\protect\citeauthoryear{{Kato}, {Nogami}, {Baba}, {Hanson}  \&
  {Poyner}}{{Kato} et~al.}{2000}]{KatoNormalOCRBOO2000MNRAS.315..140K}
{Kato} T.,  {Nogami} D.,  {Baba} H.,  {Hanson} G.,   {Poyner} G.,  2000,
  \mn@doi [\mnras] {10.1046/j.1365-8711.2000.03440.x}, \href
  {https://ui.adsabs.harvard.edu/abs/2000MNRAS.315..140K} {315, 140}

\bibitem[\protect\citeauthoryear{{Kato}, {Stubbings}, {Monard}, {Pearce}  \&
  {Nelson}}{{Kato} et~al.}{2001}]{KatoStandStillV8032001IBVS.5091....1K}
{Kato} T.,  {Stubbings} R.,  {Monard} B.,  {Pearce} A.,   {Nelson} P.,  2001,
  Information Bulletin on Variable Stars, \href
  {https://ui.adsabs.harvard.edu/abs/2001IBVS.5091....1K} {5091, 1}

\bibitem[\protect\citeauthoryear{{Kato}, {Stubbings}, {Monard}, {Butterworth},
  {Bolt}  \& {Richards}}{{Kato} et~al.}{2004}]{Kato2004PASJ...56S..89K}
{Kato} T.,  {Stubbings} R.,  {Monard} B.,  {Butterworth} N.~D.,  {Bolt} G.,
  {Richards} T.,  2004, \mn@doi [\pasj] {10.1093/pasj/56.sp1.S89}, \href
  {https://ui.adsabs.harvard.edu/abs/2004PASJ...56S..89K} {56, S89}

\bibitem[\protect\citeauthoryear{{Kato} et~al.,}{{Kato}
  et~al.}{2012}]{Kato2012PASJ...64...21K}
{Kato} T.,  et~al., 2012, \mn@doi [\pasj] {10.1093/pasj/64.1.21}, \href
  {https://ui.adsabs.harvard.edu/abs/2012PASJ...64...21K} {64, 21}

\bibitem[\protect\citeauthoryear{{Kato}, {Hambsch}  \& {Monard}}{{Kato}
  et~al.}{2015}]{Kato2015}
{Kato} T.,  {Hambsch} F.-J.,   {Monard} B.,  2015, \mn@doi [\pasj]
  {10.1093/pasj/psv010}, \href
  {https://ui.adsabs.harvard.edu/abs/2015PASJ...67L...2K} {67, L2}

\bibitem[\protect\citeauthoryear{{Kotko}, {Lasota}, {Dubus}  \&
  {Hameury}}{{Kotko} et~al.}{2012}]{Kotko2012}
{Kotko} I.,  {Lasota} J.~P.,  {Dubus} G.,   {Hameury} J.~M.,  2012, \mn@doi
  [\aap] {10.1051/0004-6361/201219156}, \href
  {https://ui.adsabs.harvard.edu/abs/2012A&A...544A..13K} {544, A13}

\bibitem[\protect\citeauthoryear{{Kukarkin} et~al.,}{{Kukarkin}
  et~al.}{1969}]{Kukarkin1969gcvs.book.....K}
{Kukarkin} B.~V.,  et~al., 1969, {General Catalogue of Variable Stars.
  Volume\_1. Constellations Andromeda - Grus.}

\bibitem[\protect\citeauthoryear{{Kupfer} et~al.,}{{Kupfer}
  et~al.}{2015}]{Kupfer2015SDSS1908}
{Kupfer} T.,  et~al., 2015, \mn@doi [\mnras] {10.1093/mnras/stv1609}, \href
  {https://ui.adsabs.harvard.edu/abs/2015MNRAS.453..483K} {453, 483}

\bibitem[\protect\citeauthoryear{{Lasota}, {Hameury}  \& {Hure}}{{Lasota}
  et~al.}{1995}]{LasotaEMTmodel1995A&A...302L..29L}
{Lasota} J.~P.,  {Hameury} J.~M.,   {Hure} J.~M.,  1995, \aap, \href
  {https://ui.adsabs.harvard.edu/abs/1995A&A...302L..29L} {302, L29}

\bibitem[\protect\citeauthoryear{{Lasota}, {Dubus}  \& {Kruk}}{{Lasota}
  et~al.}{2008}]{Lasota2008UltraCompact}
{Lasota} J.~P.,  {Dubus} G.,   {Kruk} K.,  2008, \mn@doi [\aap]
  {10.1051/0004-6361:200809658}, \href
  {https://ui.adsabs.harvard.edu/abs/2008A&A...486..523L} {486, 523}

\bibitem[\protect\citeauthoryear{{Levitan} et~al.,}{{Levitan}
  et~al.}{2011}]{Levitan2011PTF0719}
{Levitan} D.,  et~al., 2011, \mn@doi [\apj] {10.1088/0004-637X/739/2/68}, \href
  {https://ui.adsabs.harvard.edu/abs/2011ApJ...739...68L} {739, 68}

\bibitem[\protect\citeauthoryear{{Levitan} et~al.,}{{Levitan}
  et~al.}{2013}]{Levitan2013PTF1}
{Levitan} D.,  et~al., 2013, \mn@doi [\mnras] {10.1093/mnras/sts672}, \href
  {https://ui.adsabs.harvard.edu/abs/2013MNRAS.430..996L} {430, 996}

\bibitem[\protect\citeauthoryear{{Levitan}, {Groot}, {Prince}, {Kulkarni},
  {Laher}, {Ofek}, {Sesar}  \& {Surace}}{{Levitan}
  et~al.}{2015}]{Levitan2015longterm}
{Levitan} D.,  {Groot} P.~J.,  {Prince} T.~A.,  {Kulkarni} S.~R.,  {Laher} R.,
  {Ofek} E.~O.,  {Sesar} B.,   {Surace} J.,  2015, \mn@doi [\mnras]
  {10.1093/mnras/stu2105}, \href
  {https://ui.adsabs.harvard.edu/abs/2015MNRAS.446..391L} {446, 391}

\bibitem[\protect\citeauthoryear{{Lightkurve Collaboration}
  et~al.,}{{Lightkurve Collaboration} et~al.}{2018}]{lcsoftware2018}
{Lightkurve Collaboration} et~al., 2018, {Lightkurve: Kepler and TESS time
  series analysis in Python}, Astrophysics Source Code Library (\mn@eprint
  {ascl} {1812.013})

\bibitem[\protect\citeauthoryear{{Linnell Nemec} \& {Nemec}}{{Linnell Nemec} \&
  {Nemec}}{1985}]{Nemec1985}
{Linnell Nemec} A.~F.,  {Nemec} J.~M.,  1985, \mn@doi [\aj] {10.1086/113936},
  \href {https://ui.adsabs.harvard.edu/abs/1985AJ.....90.2317L} {90, 2317}

\bibitem[\protect\citeauthoryear{{Lomb}}{{Lomb}}{1976}]{Lomb1976}
{Lomb} N.~R.,  1976, \mn@doi [\apss] {10.1007/BF00648343}, \href
  {https://ui.adsabs.harvard.edu/abs/1976Ap&SS..39..447L} {39, 447}

\bibitem[\protect\citeauthoryear{{Masci} et~al.,}{{Masci}
  et~al.}{2019a}]{ZTF2019Masci}
{Masci} F.~J.,  et~al., 2019a, \mn@doi [\pasp] {10.1088/1538-3873/aae8ac},
  \href {https://ui.adsabs.harvard.edu/abs/2019PASP..131a8003M} {131, 018003}

\bibitem[\protect\citeauthoryear{{Masci} et~al.,}{{Masci}
  et~al.}{2019b}]{MasciZTF2019}
{Masci} F.~J.,  et~al., 2019b, \mn@doi [\pasp] {10.1088/1538-3873/aae8ac},
  \href {https://ui.adsabs.harvard.edu/abs/2019PASP..131a8003M} {131, 018003}

\bibitem[\protect\citeauthoryear{{Menou}, {Perna}  \& {Hernquist}}{{Menou}
  et~al.}{2002}]{Menou2002}
{Menou} K.,  {Perna} R.,   {Hernquist} L.,  2002, \mn@doi [\apjl]
  {10.1086/338909}, \href
  {https://ui.adsabs.harvard.edu/abs/2002ApJ...564L..81M} {564, L81}

\bibitem[\protect\citeauthoryear{{Meyer} \& {Meyer-Hofmeister}}{{Meyer} \&
  {Meyer-Hofmeister}}{1981}]{Meyer1981DIM}
{Meyer} F.,  {Meyer-Hofmeister} E.,  1981, \aap, \href
  {https://ui.adsabs.harvard.edu/abs/1981A&A...104L..10M} {104, L10}

\bibitem[\protect\citeauthoryear{{Motsoaledi} et~al.,}{{Motsoaledi}
  et~al.}{2021}]{ASASSN21brperiod}
{Motsoaledi} M.,  et~al., 2021, The Astronomer's Telegram, \href
  {https://ui.adsabs.harvard.edu/abs/2021ATel14421....1M} {14421, 1}

\bibitem[\protect\citeauthoryear{{Nelemans}, {Yungelson}  \& {Portegies
  Zwart}}{{Nelemans} et~al.}{2004}]{Nelemans2004lisa}
{Nelemans} G.,  {Yungelson} L.~R.,   {Portegies Zwart} S.~F.,  2004, \mn@doi
  [\mnras] {10.1111/j.1365-2966.2004.07479.x}, \href
  {https://ui.adsabs.harvard.edu/abs/2004MNRAS.349..181N} {349, 181}

\bibitem[\protect\citeauthoryear{{O'Donoghue}, {Menzies}  \&
  {Hill}}{{O'Donoghue} et~al.}{1987}]{V8031987MNRAS.227..347O}
{O'Donoghue} D.,  {Menzies} J.~W.,   {Hill} P.~W.,  1987, \mn@doi [\mnras]
  {10.1093/mnras/227.2.347}, \href
  {https://ui.adsabs.harvard.edu/abs/1987MNRAS.227..347O} {227, 347}

\bibitem[\protect\citeauthoryear{{Osaki}}{{Osaki}}{1989}]{Osaki1989SOModelPASJ...41.1005O}
{Osaki} Y.,  1989, \pasj, \href
  {https://ui.adsabs.harvard.edu/abs/1989PASJ...41.1005O} {41, 1005}

\bibitem[\protect\citeauthoryear{{Osaki} \& {Kato}}{{Osaki} \&
  {Kato}}{2013}]{Osaki2013KeplerV15}
{Osaki} Y.,  {Kato} T.,  2013, \mn@doi [\pasj] {10.1093/pasj/65.3.50}, \href
  {https://ui.adsabs.harvard.edu/abs/2013PASJ...65...50O} {65, 50}

\bibitem[\protect\citeauthoryear{{Osaki} \& {Meyer}}{{Osaki} \&
  {Meyer}}{2003}]{OsakiMeyer2003A&A...401..325O}
{Osaki} Y.,  {Meyer} F.,  2003, \mn@doi [\aap] {10.1051/0004-6361:20030115},
  \href {https://ui.adsabs.harvard.edu/abs/2003A&A...401..325O} {401, 325}

\bibitem[\protect\citeauthoryear{{Patterson} et~al.,}{{Patterson}
  et~al.}{1998}]{Patterson1998EG_Candri}
{Patterson} J.,  et~al., 1998, \mn@doi [\pasp] {10.1086/316252}, \href
  {https://ui.adsabs.harvard.edu/abs/1998PASP..110.1290P} {110, 1290}

\bibitem[\protect\citeauthoryear{{Patterson}, {Walker}, {Kemp}, {O'Donoghue},
  {Bos}  \& {Stubbings}}{{Patterson}
  et~al.}{2000}]{V803Patterson2000PASP..112..625P}
{Patterson} J.,  {Walker} S.,  {Kemp} J.,  {O'Donoghue} D.,  {Bos} M.,
  {Stubbings} R.,  2000, \mn@doi [\pasp] {10.1086/316561}, \href
  {https://ui.adsabs.harvard.edu/abs/2000PASP..112..625P} {112, 625}

\bibitem[\protect\citeauthoryear{{Patterson} et~al.,}{{Patterson}
  et~al.}{2002}]{Patterson2002WZSGe}
{Patterson} J.,  et~al., 2002, \mn@doi [\pasp] {10.1086/341696}, \href
  {https://ui.adsabs.harvard.edu/abs/2002PASP..114..721P} {114, 721}

\bibitem[\protect\citeauthoryear{{Ramsay} et~al.,}{{Ramsay}
  et~al.}{2010}]{RamsayKLDra2010MNRAS.407.1819R}
{Ramsay} G.,  et~al., 2010, \mn@doi [\mnras]
  {10.1111/j.1365-2966.2010.17019.x}, \href
  {https://ui.adsabs.harvard.edu/abs/2010MNRAS.407.1819R} {407, 1819}

\bibitem[\protect\citeauthoryear{{Ramsay} et~al.,}{{Ramsay}
  et~al.}{2018}]{RamsayCatalog}
{Ramsay} G.,  et~al., 2018, \mn@doi [\aap] {10.1051/0004-6361/201834261}, \href
  {https://ui.adsabs.harvard.edu/abs/2018A&A...620A.141R} {620, A141}

\bibitem[\protect\citeauthoryear{{Ricker} et~al.,}{{Ricker}
  et~al.}{2014}]{TESS2014SPIE}
{Ricker} G.~R.,  et~al., 2014, in {Oschmann} Jacobus~M. J.,  {Clampin} M.,
  {Fazio} G.~G.,   {MacEwen} H.~A.,  eds,  Society of Photo-Optical
  Instrumentation Engineers (SPIE) Conference Series Vol. 9143, Space
  Telescopes and Instrumentation 2014: Optical, Infrared, and Millimeter Wave.
  p. 914320 (\mn@eprint {arXiv} {1406.0151}), \mn@doi{10.1117/12.2063489}

\bibitem[\protect\citeauthoryear{{Rivera Sandoval} \& {Maccarone}}{{Rivera
  Sandoval} \& {Maccarone}}{2019}]{2019RSM}
{Rivera Sandoval} L.~E.,  {Maccarone} T.~J.,  2019, \mn@doi [\mnras]
  {10.1093/mnrasl/sly205}, \href
  {https://ui.adsabs.harvard.edu/abs/2019MNRAS.483L...6R} {483, L6}

\bibitem[\protect\citeauthoryear{{Rivera Sandoval}, {Maccarone}, {Cavecchi},
  {Britt}  \& {Zurek}}{{Rivera Sandoval}
  et~al.}{2020a}]{RiveraSandoval60minOutburst2020arXiv201210356R}
{Rivera Sandoval} L.~E.,  {Maccarone} T.~J.,  {Cavecchi} Y.,  {Britt} C.,
  {Zurek} D.,  2020a, arXiv e-prints, \href
  {https://ui.adsabs.harvard.edu/abs/2020arXiv201210356R} {p. arXiv:2012.10356}

\bibitem[\protect\citeauthoryear{{Rivera Sandoval}, {Maccarone}  \& {Pichardo
  Marcano}}{{Rivera Sandoval}
  et~al.}{2020b}]{RiveraSandovalYearlong2020ApJ...900L..37R}
{Rivera Sandoval} L.~E.,  {Maccarone} T.~J.,   {Pichardo Marcano} M.,  2020b,
  \mn@doi [\apjl] {10.3847/2041-8213/abb130}, \href
  {https://ui.adsabs.harvard.edu/abs/2020ApJ...900L..37R} {900, L37}

\bibitem[\protect\citeauthoryear{{Roelofs}, {Groot}, {Marsh}, {Steeghs},
  {Barros}  \& {Nelemans}}{{Roelofs} et~al.}{2005}]{RoelofsSDSS}
{Roelofs} G.~H.~A.,  {Groot} P.~J.,  {Marsh} T.~R.,  {Steeghs} D.,  {Barros}
  S.~C.~C.,   {Nelemans} G.,  2005, \mn@doi [\mnras]
  {10.1111/j.1365-2966.2005.09186.x}, \href
  {https://ui.adsabs.harvard.edu/abs/2005MNRAS.361..487R} {361, 487}

\bibitem[\protect\citeauthoryear{{Roelofs}, {Groot}, {Nelemans}, {Marsh}  \&
  {Steeghs}}{{Roelofs} et~al.}{2007}]{V803Roelofs2007MNRAS.379..176R}
{Roelofs} G.~H.~A.,  {Groot} P.~J.,  {Nelemans} G.,  {Marsh} T.~R.,   {Steeghs}
  D.,  2007, \mn@doi [\mnras] {10.1111/j.1365-2966.2007.11931.x}, \href
  {https://ui.adsabs.harvard.edu/abs/2007MNRAS.379..176R} {379, 176}

\bibitem[\protect\citeauthoryear{{Scargle}}{{Scargle}}{1982}]{Scargle1982}
{Scargle} J.~D.,  1982, \mn@doi [\apj] {10.1086/160554}, \href
  {https://ui.adsabs.harvard.edu/abs/1982ApJ...263..835S} {263, 835}

\bibitem[\protect\citeauthoryear{{Schwartz}}{{Schwartz}}{1998}]{SchwartzSNKlDRa1998IAUC.6982....1S}
{Schwartz} M.,  1998, \iaucirc, \href
  {https://ui.adsabs.harvard.edu/abs/1998IAUC.6982....1S} {6982, 1}

\bibitem[\protect\citeauthoryear{{Schwarzenberg-Czerny}}{{Schwarzenberg-Czerny}}{1997}]{pdmbeta97}
{Schwarzenberg-Czerny} A.,  1997, \mn@doi [\apj] {10.1086/304832}, \href
  {https://ui.adsabs.harvard.edu/abs/1997ApJ...489..941S} {489, 941}

\bibitem[\protect\citeauthoryear{{Schwarzenberg-Czerny}}{{Schwarzenberg-Czerny}}{2003}]{GuidePeriodSearch2003}
{Schwarzenberg-Czerny} A.,  2003, in {Sterken} C.,  ed.,  Astronomical Society
  of the Pacific Conference Series Vol. 292, Interplay of Periodic, Cyclic and
  Stochastic Variability in Selected Areas of the H-R Diagram. p.~383

\bibitem[\protect\citeauthoryear{{Shappee} et~al.,}{{Shappee}
  et~al.}{2014}]{ASASSN2014}
{Shappee} B.~J.,  et~al., 2014, \mn@doi [\apj] {10.1088/0004-637X/788/1/48},
  \href {https://ui.adsabs.harvard.edu/abs/2014ApJ...788...48S} {788, 48}

\bibitem[\protect\citeauthoryear{{Smak}}{{Smak}}{1982}]{Smak1982}
{Smak} J.,  1982, \actaa, \href
  {https://ui.adsabs.harvard.edu/abs/1982AcA....32..199S} {32, 199}

\bibitem[\protect\citeauthoryear{{Smak}}{{Smak}}{1983}]{Smak1983AcA....33..333SHelium}
{Smak} J.,  1983, \actaa, \href
  {https://ui.adsabs.harvard.edu/abs/1983AcA....33..333S} {33, 333}

\bibitem[\protect\citeauthoryear{{Smak}}{{Smak}}{1999}]{Smak1999AcA....49..391S}
{Smak} J.,  1999, \actaa, \href
  {https://ui.adsabs.harvard.edu/abs/1999AcA....49..391S} {49, 391}

\bibitem[\protect\citeauthoryear{{Smak}}{{Smak}}{2009a}]{Smak2009AcA....59...89S}
{Smak} J.,  2009a, \actaa, \href
  {https://ui.adsabs.harvard.edu/abs/2009AcA....59...89S} {59, 89}

\bibitem[\protect\citeauthoryear{{Smak}}{{Smak}}{2009b}]{Smak2009AcA....59..103S}
{Smak} J.,  2009b, \actaa, \href
  {https://ui.adsabs.harvard.edu/abs/2009AcA....59..103S} {59, 103}

\bibitem[\protect\citeauthoryear{{Smak}}{{Smak}}{2009c}]{Smak2009AcA....59..121S}
{Smak} J.,  2009c, \actaa, \href
  {https://ui.adsabs.harvard.edu/abs/2009AcA....59..121S} {59, 121}

\bibitem[\protect\citeauthoryear{{Smak}}{{Smak}}{2009d}]{Smak2009AcA....59..419S}
{Smak} J.,  2009d, \actaa, \href
  {https://ui.adsabs.harvard.edu/abs/2009AcA....59..419S} {59, 419}

\bibitem[\protect\citeauthoryear{{Smith} et~al.,}{{Smith}
  et~al.}{2012}]{KeplerLCMethod2012}
{Smith} J.~C.,  et~al., 2012, \mn@doi [\pasp] {10.1086/667697}, \href
  {https://ui.adsabs.harvard.edu/abs/2012PASP..124.1000S} {124, 1000}

\bibitem[\protect\citeauthoryear{{Solheim}}{{Solheim}}{2010}]{Solheim2010}
{Solheim} J.~E.,  2010, \mn@doi [\pasp] {10.1086/656680}, \href
  {https://ui.adsabs.harvard.edu/abs/2010PASP..122.1133S} {122, 1133}

\bibitem[\protect\citeauthoryear{{Stellingwerf}}{{Stellingwerf}}{1978}]{Stellingwerf1978PDM}
{Stellingwerf} R.~F.,  1978, \mn@doi [\apj] {10.1086/156444}, \href
  {https://ui.adsabs.harvard.edu/abs/1978ApJ...224..953S} {224, 953}

\bibitem[\protect\citeauthoryear{{Sunny Wong}, {van Roestel}, {Kupfer}  \&
  {Bildsten}}{{Sunny Wong} et~al.}{2021}]{Wong2021RNAAS...5....3S}
{Sunny Wong} T.~L.,  {van Roestel} J.,  {Kupfer} T.,   {Bildsten} L.,  2021,
  \mn@doi [Research Notes of the American Astronomical Society]
  {10.3847/2515-5172/abd7fa}, \href
  {https://ui.adsabs.harvard.edu/abs/2021RNAAS...5....3S} {5, 3}

\bibitem[\protect\citeauthoryear{{Szkody} et~al.,}{{Szkody}
  et~al.}{2020}]{Szkody2020AJ....159..198S}
{Szkody} P.,  et~al., 2020, \mn@doi [\aj] {10.3847/1538-3881/ab7cce}, \href
  {https://ui.adsabs.harvard.edu/abs/2020AJ....159..198S} {159, 198}

\bibitem[\protect\citeauthoryear{{Tonry} et~al.,}{{Tonry}
  et~al.}{2018}]{Atlas2018PASP..130f4505T}
{Tonry} J.~L.,  et~al., 2018, \mn@doi [\pasp] {10.1088/1538-3873/aabadf}, \href
  {https://ui.adsabs.harvard.edu/abs/2018PASP..130f4505T} {130, 064505}

\bibitem[\protect\citeauthoryear{{Tsugawa} \& {Osaki}}{{Tsugawa} \&
  {Osaki}}{1997}]{Tsugawa1997PASJ...49...75T}
{Tsugawa} M.,  {Osaki} Y.,  1997, \mn@doi [\pasj] {10.1093/pasj/49.1.75}, \href
  {https://ui.adsabs.harvard.edu/abs/1997PASJ...49...75T} {49, 75}

\bibitem[\protect\citeauthoryear{{Uemura} et~al.,}{{Uemura}
  et~al.}{2005}]{UemuraNoprecurso2005A&A...432..261U}
{Uemura} M.,  et~al., 2005, \mn@doi [\aap] {10.1051/0004-6361:20042004}, \href
  {https://ui.adsabs.harvard.edu/abs/2005A&A...432..261U} {432, 261}

\bibitem[\protect\citeauthoryear{{Vanderspek}}{{Vanderspek}}{2019}]{TESShandbook2019ESS.....433312V}
{Vanderspek} R.,  2019, in AAS/Division for Extreme Solar Systems Abstracts. p.
  333.12

\bibitem[\protect\citeauthoryear{Virtanen et~al.,}{Virtanen
  et~al.}{2020}]{SciPy-NMeth}
Virtanen P.,  et~al., 2020, \mn@doi [Nature Methods]
  {10.1038/s41592-019-0686-2}, \href {https://rdcu.be/b08Wh} {17, 261}

\bibitem[\protect\citeauthoryear{{Warner}}{{Warner}}{1995}]{Warner1995SOSUUMAAp&SS.226..187W}
{Warner} B.,  1995, \mn@doi [\apss] {10.1007/BF00627371}, \href
  {https://ui.adsabs.harvard.edu/abs/1995Ap&SS.226..187W} {226, 187}

\bibitem[\protect\citeauthoryear{{Wood}, {Casey}, {Garnavich}  \&
  {Haag}}{{Wood} et~al.}{2002}]{WoodKlDraSumperHump2002MNRAS.334...87W}
{Wood} M.~A.,  {Casey} M.~J.,  {Garnavich} P.~M.,   {Haag} B.,  2002, \mn@doi
  [\mnras] {10.1046/j.1365-8711.2002.05484.x}, \href
  {https://ui.adsabs.harvard.edu/abs/2002MNRAS.334...87W} {334, 87}

\bibitem[\protect\citeauthoryear{{Wyrzykowski}, {Hodgkin}, {Blogorodnova},
  {Koposov}  \& {Burgon}}{{Wyrzykowski} et~al.}{2012}]{GaiaAlert2012}
{Wyrzykowski} {\L}.,  {Hodgkin} S.,  {Blogorodnova} N.,  {Koposov} S.,
  {Burgon} R.,  2012, in 2nd Gaia Follow-up Network for Solar System Objects.
  p.~21 (\mn@eprint {arXiv} {1210.5007})

\bibitem[\protect\citeauthoryear{{York} et~al.,}{{York}
  et~al.}{2000}]{2000AJ....120.1579Y}
{York} D.~G.,  et~al., 2000, \mn@doi [\aj] {10.1086/301513}, \href
  {https://ui.adsabs.harvard.edu/abs/2000AJ....120.1579Y} {120, 1579}

\bibitem[\protect\citeauthoryear{{van Roestel} et~al.,}{{van Roestel}
  et~al.}{2021}]{vanroestel2021}
{van Roestel} J.,  et~al., 2021, arXiv e-prints, \href
  {https://ui.adsabs.harvard.edu/abs/2021arXiv210502261V} {p. arXiv:2105.02261}

\makeatother
\end{thebibliography}




\FloatBarrier


\appendix

\section*{Appendix}
\label{sec:appendix}

\section{Sources without detected superoutbursts}
\label{sec:NoSO}

\begin{table*}
\centering
\caption{The AM~CVn in our sample for which no outbursts were detected with \emph{TESS}. Periods are taken from \citet[][]{RamsayCatalog} except below the double line.
  Periods with '(sh)' next to the numbers are those estimated from superhumps, and (p) denotes systems where the period is estimated based on outburst properties. In some cases, the non-detections are unsurprising because the expected recurrence times are less than once per sector, and sometimes considerably less, but in other cases, the non-detections are unsurprising because past outbursts did not reach \emph{TESS} detection limits. The magnitude ranges are taken from \citet[][]{RamsayCatalog} and the ZTF public survey data.}
 \begin{adjustbox}{totalheight=\textheight-5\baselineskip}
\begin{tabular}{|c|c|c|c|c|c|c|} 
\hline
Source   & RA (J2000) & DEC (J2000) &\emph{TESS} sectors   & $P_{orb}$ (min)& Mag. Range (filter) &  Comments    \\ 
\hline
V407 Vul  &19:14:26.09& +24:56:44.6      & 14 & 9.5  & 19.9 (V)  &direct impact accretor \\ 
\hline
ES Cet    &   02:00:52.2360	&-09:24:31.645	   & 30      &  10.4 & 16.5–16.8       & direct impact accretor  \\
\hline
AM CVn    &12:34:54.6230 & +37:37:44.120      & 22      &      17.1 & 14.2  &    persistent  \\ 
\hline
PTF1J1919+4815 &19:19:05.1825 & +48:15:06.031 & 14-15   &  22.5   & 18.2–21.8     &  outbursts peak at $\sim18$  \\ 
\hline
YZ LMi &09:26:38.7243 & +36:24:02.463         & 21             &   28.3 & 16.6–19.6 &      \\ 
\hline
CRTS J0910-2008 &09:10:17.4492 & -20:08:12.462 & 8      &29.7 (sh) & 14.0–20.4 (g)       &         \\ 
\hline
CRTS J0105+1903 &01:05:50.0988 & +19:03:17.000 & 17             &   31.6  & 16.3-19.6     &   \\ 
\hline
PTF1J1632+3511 & 16:32:39.39  & +35:11:07.3  & 24-25          &     32.7(p) & 17.9–23.0    &  peak at $V\sim18$  \\ 
\hline
CRTS J0744+3254 &07:44:19.7435 & +32:54:48.242 & 20& 33(p)         & 17.4–21.1           &   \\ 
\hline
V406 Hya  &09:05:54.7498& -05:36:08.482      & 8      & 33.8 & 14.5–19.7   &   \\ 
\hline
PTF1 J0435+0029 &04:35:17.73& +00:29:40.7	 & 5, 32 &   34.3    & 18.4 (R) – 22.3 (g)  &   peak at $R$=18.4 \\ 
\hline
SDSS J1730+5545 &17:30:47.5876& +55:45:18.463 & 14-26          &   35.2 & 18.5 (V) –20.1     & peaks at $V=18.5$ \\ 
\hline
NSV 1440 &03:55:17.9490& -82:26:11.305       & 1,12-13, 27-28 &  37.5 (sh)  & 12.4(V) – 17.9(G)       & $V$=12.4 in outburst\\ 
\hline
V744 And  &01:29:40.0646& +38:42:10.442      & 17   &   37.6     & 14.5–19.8    &     \\ 
\hline
SDSSJ1721+2733 &17:21:02.4797& +27:33:01.236  & 25-26 &38.1     & 16.0–20.1     &          \\ 
\hline
ASASSN-14ei &02:55:33.2826& -47:50:42.318    & 3-4, 30        &  43(sh)  & 11.9–17.6 (B)      &   \\ 
\hline
SDSSJ1525+3600 &15:25:09.5669& +36:00:54.683  & 24   &44.3 & 20.2            &             \\ 
\hline
SDSSJ1411+4812 &14:11:18.3211& +48:12:57.515   & 16,22-23       &  46.0   & 19.4–19.7     &   \\ 
\hline
GP Com &13:05:42.4008& +18:01:03.765          & 23             &  46.6   & 15.9–16.3     &   \\ 
\hline
CRTS J0450-0931 &04:50:20.0	& -09:31:13  & 5,32           &    47.3 (sh)  & 14.8–20.5    &   \\ 
\hline
SDSS J0902+3819 &09:02:21.3623& +38:19:41.841  & 21             &   48.3  & 13.8 (V) – 20.2 (g)     &   \\ 
\hline
Gaia14aae  &16:11:33.9745& +63:08:31.886      & 14-26          &   49.7  & 13.6 (V) – 18.7 (g)     &   \\ 
\hline
ASASSN-17fp &18:08:51.10& -73:04:04.2     & 12-13          &   51.0 (sh)  & 15.7–20+     &   \\ 
\hline
SDSS J1208+3550 &12:08:42.0093& +35:50:25.391  &    22            &  53.0  & 18.9–19.4      &   \\ 
\hline
SDSS J1642+1934 &16:42:28.0669& +19:34:10.124  & 25             &  54.2 & 20.3    &   \\ 
\hline
SDSS J1552+3201 &15:52:52.4801& +32:01:51.018  & 24             &  56.3  & 20.2–20.6      &   \\ 
\hline
SDSS J1137+4054 &11 37 32.3218& +40:54:58.496  & 22             &  59.6   & 19.0     &   \\ 
\hline
V396 Hya &13:12:46.4403& -23:21:32.601        & 10             &      65.1 & 17.6    &   \\ 
\hline
SDSS J1319+5915 &13:19:54.5178& +59:15:14.563  & 15-16,22       &  65.6  & 19.1  &   \\ 
\hline
CRTS J0844-0128 &08:44:13.6& -01:28:07 & 8              &  & 17.4-20.3        &   \\ 
\hline
PTF1 J0857+0729 &08:57:24.27& +07:29:46.7	  & 8              &      & 18.6-21.8    &   \\ 
\hline
ASASSN-14fv &23:29:55.13& +44:56:14.4    & 16-17          &      & 14.6 (V) – 20.5 (B)    &   \\ 
\hline
\hline 
ASASSN-21br &16:18:10.44& -51:54:15.8   &  12 & 38.6                & >17.1-13.6      &  period from \citet[][]{ASASSN21brperiod} \\ 
\hline
ASASSN-21au &14:23:52.820& +78:30:13.370    & 12             &  58.4 (sh)   & 20.6-13.4 (g)     &  Average period and g-mag range from \citep{2021RSAU} \\ 
\hline
ZTF J1905+3134 &19:05:11.34& +31:34:32.37  & 14 &17.2           &   20.90 (g) &            persistent; period from \citep[][]{Burdge2020ApJ} \\ 
\hline
ZTF J2228+4949 &22:28:27.07& +49:49:16.44  & 16-17& 28.6    
& 19.2758(G)&           persistent, $\sim 19$th mag; period from \citep[][]{Burdge2020ApJ} \\ 
\hline
ZTF18acgmwpt &07:01:15.85& +50:23:21.5  &  20  &  & 17.5 -20.50 (g)             &   from ZTF g        \\ 
\hline
ZTF18acnnabo &08:20:47.6& 68:04:24.0  &20, 26 & & 16.5 -19.8 (g) & from ZTF \\
\hline
ZTF18acujsfl &04:49:30.1& -02:51:53.7  & 5,32 & & 15.7 -20.3(g) & from ZTF \\
\hline
ZTF19abzzuin &08:44:19.7& +06:39:50.2  & 7, 34 & & 18.1 -21.16(g) & from ZTF \\
\hline
ZTF18aavetqn &19:18:42.0& +44:49:12.3  & 14,15&  & 14.9 -19.9(g)& from ZTF \\
\hline
\label{nondetect}
\end{tabular}
\end{adjustbox}
\end{table*}

\FloatBarrier

\section{ASASSN-14cc Periodogram}

 \begin{figure*}[h]
	\includegraphics[width= 1.0 \columnwidth]{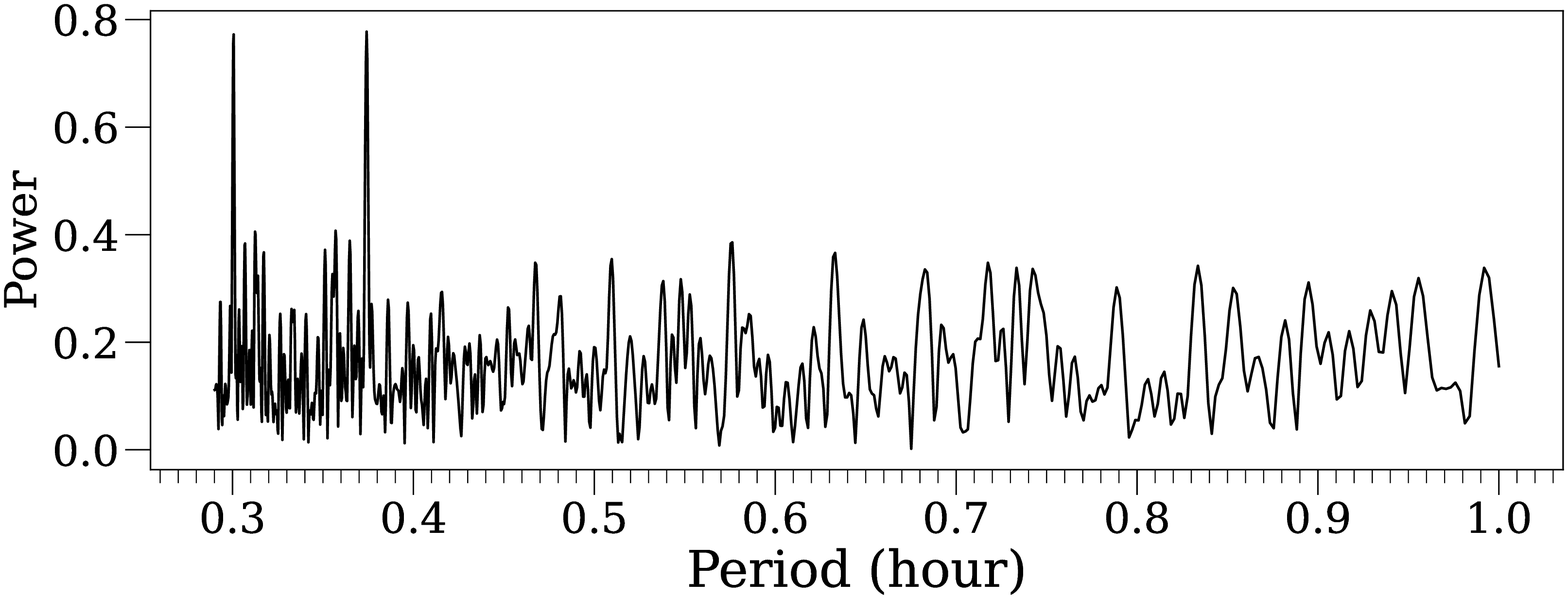}
    \caption{Lomb-Scargle ASASSN-14cc in Sector 27. The periodogram shows the superhump period at 22.5 minutes, and another peak at 18.1 minutes which is the beat period between the superhump period and the \emph{TESS} sampling of 10 minutes.}
    \label{fig:LombScargle14ccappendix}
\end{figure*}


\bsp	
\label{lastpage}
\end{document}